\documentclass[11pt,a4paper]{article}
\usepackage{jcappub}
\newcommand{\ukarcmin}{\ensuremath{\mu{\rm K{-}arcmin}}}

\newcommand{\mvar}{\ensuremath{M_{200}}}

\newcommand{\comment}[1]{}

\usepackage{hyperref}
\usepackage{xcolor}
\usepackage[normalem]{ulem}
\usepackage{graphicx}
\usepackage{amsmath}

\usepackage{dcolumn}
\usepackage{multirow}
\usepackage{tabularx}
\newcolumntype{C}[1]{>{\centering\let\newline\\\arraybackslash\hspace{0pt}}m{#1}}

\newcolumntype{L}{D{.}{.}{2,4}}
\title{Measuring galaxy cluster masses with CMB lensing using a Maximum Likelihood estimator: Statistical and systematic error budgets for future experiments}

\author[a]{Srinivasan Raghunathan,}%\email{srinivasan.raghunathan@unimelb.edu.au}
\author[a]{Sanjaykumar Patil,}
\author[b]{Eric J. Baxter,}
\author[a]{Federico Bianchini,}
\author[c, d]{Lindsey E. Bleem,}
\author[d, e]{Thomas M. Crawford,}
\author[f]{Gilbert P. Holder,}
\author[e]{Alessandro Manzotti,}
\author[a]{and Christian L. Reichardt}
\affiliation[a]{School of Physics, University of Melbourne, Parkville VIC 3010, Australia}
\affiliation[b]{Department of Physics and Astronomy, University of Pennsylvania, Philadelphia, PA 19104, USA}
\affiliation[c]{Argonne National Laboratory, High-Energy Physics Division, 9700 S. Cass Avenue, Argonne, IL, USA 60439}
\affiliation[d]{Kavli Institute for Cosmological Physics, University of Chicago, 5640 South Ellis Avenue, Chicago, IL 60637, USA}
\affiliation[e]{Department of Astronomy and Astrophysics, University of Chicago, 5640 South Ellis Avenue, Chicago, IL 60637, USA}
\affiliation[f]{Department of Astronomy and Department of Physics, University of Illinois, 1002 West Green St., Urbana, IL 61801}
\emailAdd{srinivasan.raghunathan@unimelb.edu.au}

\abstract{We develop a Maximum Likelihood estimator (MLE) to measure the masses of galaxy clusters through the impact of gravitational lensing on the temperature and polarization anisotropies of the cosmic microwave background (CMB). We show that, at low noise levels in temperature,  this optimal estimator outperforms the standard quadratic estimator by a factor of two. For polarization, we show that the Stokes Q/U maps can be used instead of the traditional E- and B-mode maps without losing information. We test and quantify the bias in the recovered lensing mass for a comprehensive list of potential systematic errors. Using realistic simulations, we examine  the cluster mass uncertainties from CMB-cluster lensing as a function of an experiment's beam size and noise level. We predict the cluster mass uncertainties will be 3 - 6\% for SPT-3G, AdvACT, and Simons Array experiments with 10,000 clusters and less than 1\% for the CMB-S4 experiment with a sample containing 100,000 clusters. The mass constraints from CMB polarization are very sensitive to the experimental beam size and map noise level: for a factor of three reduction in either the beam size or noise level, the lensing signal-to-noise improves by roughly a factor of two.}

\begin{document}
\maketitle

\section{Introduction}
As the largest collapsed halos, galaxy clusters are important probes of cosmology and structure growth in the Universe. 
Measuring the number density of galaxy clusters as a function of mass and redshift has the potential to yield some of the tightest constraints on the dark energy equation of state, especially at high redshifts ($z > 1$) \citep{allen2011}. Current CMB, optical, and X-ray surveys have begun to return large samples of galaxy clusters \citep{xray_clusters_mcxc_2011, rykoff2014, rykoff2016, bleem2015, ACTSZ2013, PLANCKSZ2016}, and these samples are expected to rise substantially to over 100,000 with upcoming surveys like CMB-S4, eRosita, Euclid, or LSST \citep{erosita_science_book, lsst_science_book, euclid_science_book, cmbs4_2016}. However, while we have demonstrated the ability to find these objects, accurately determining their masses remains challenging --- and accurate mass estimates are crucial to the cosmological constraints. A mass-observable scaling relation is normally required to convert the observed quantity into cluster mass. The current uncertainties on the calibration of these scaling relations is the prime obstacle in using galaxy clusters as powerful probes of cosmology \citep{linden2014, wtgIII2014}.

Weak gravitational lensing is the primary tool to improve the mass calibration of galaxy cluster scaling relations. Weak gravitational lensing directly probes the total matter content of a galaxy cluster and can provide an unbiased measurement of the total mass. The lensed `source' can be galaxies, or  the CMB, with the CMB being the focus of this work. The CMB is particularly useful for high redshift clusters due to the difficulties in observing galaxies with sufficient signal-to-noise (SNR) in other frequencies (e.g., optical, X-ray) at very high redshifts. The CMB also has simpler systematics uncertainties, as we have extremely good measurements of the statistical properties of the CMB and we know the precise redshift where the CMB photons originate. However, the SNR of the CMB-cluster lensing signal is comparatively weak for an individual galaxy cluster, which necessitates a stacking analysis to recover the `average' mass of a cluster sample \citep{hu2007}.

\subsection{CMB-cluster lensing}

Massive galaxy clusters bend the path of CMB photons, imprinting a gravitational lensing signal on the CMB. CMB-cluster lensing refers to the gravitational lensing of the CMB by massive galaxy clusters. The scales of interest are typically a few arc-minutes, corresponding to the angular size of galaxy clusters. On these scales, the CMB is well approximated as a gradient field across the position of the galaxy cluster. The direction of the CMB gradient is nearly independent (only slightly correlated) for temperature and polarization. Gravitationally lensing of this gradient by a galaxy cluster produces a dipole-like pattern oriented with the gradient \citep{seljak2000,lewis2006}, but with the hot and cold directions swapped (\emph{cf.} see Fig. 1 of \citealt{lewis2006}). For a given cluster mass and redshift, the magnitude of this dipole scales linearly with the magnitude of the CMB gradient. Note that this implies the polarized lensing signal is an order of magnitude smaller than the temperature signal as the CMB is only $\mathcal{O}(10\%)$ polarized. For example, the lensing induced distortion due to a $2 \times 10^{14} \ M_{\odot}$ mass galaxy cluster  is $\sim 5.0$ and $0.5 \ \mu K$ in temperature and polarization respectively.

Lensing remaps the unlensed CMB to new positions based on the deflection angle of the lens. Thus, the unlensed temperature $\widetilde{T}(\hat{\textbf{n}})$ and polarization $\widetilde{Q}(\hat{\textbf{n}})$ and $\widetilde{U}(\hat{\textbf{n}})$ fields are remapped to new positions as
\begin{eqnarray}
T(\hat{\textbf{n}})& = & \widetilde{T} (\hat{\textbf{n}} + \vec{\alpha}(\hat{\textbf{n}}))\\
Q(\hat{\textbf{n}}) & = & \widetilde{Q} (\hat{\textbf{n}} + \vec{\alpha}(\hat{\textbf{n}}))\\
U(\hat{\textbf{n}}) & = & \widetilde{U} (\hat{\textbf{n}} + \vec{\alpha}(\hat{\textbf{n}}))
\end{eqnarray} where the deflection angle vector $\vec{\alpha}(\hat{\textbf{n}}) = \nabla \phi(\hat{\textbf{n}})$ is the gradient of the %underlying gravitational 
lensing potential $\phi$. 
The deflection angle depends on the mass, distance, and the profile of the lens (in this case, a galaxy cluster). 

Several methods have been proposed to extract this cluster lensing signal \emph{viz.} a Maximum Likelihood estimator (MLE) by fitting lensing templates to the lensed CMB maps \citep{lewis2006, dodelson2004, baxter2015}; a quadratic estimator (QE) which uses the correlation between the unlensed CMB gradient and the lensing signal \citep{maturi2005, hu2007, yoo2008, yoo2010, melin2015}; and a Wiener filter approach to estimate the lensing distortion \citep{holder2004}. This work is based on the MLE, which will be described in more detail in Section~\ref{sec_MLE_method}. On the data side, the CMB lensing signal due to dark matter haloes has been detected at $\sim 3 - 5 \sigma$ levels using data from ACTPol, SPT-SZ, and the Planck satellite. \citet{baxter2015} used a MLE approach to measure the lensing signal due to 513 SZ selected ($M_{_{200}} \sim 6 \times 10^{14}\ M_{\odot}$) galaxy clusters from the $2500\ \rm deg^{2}$ SPT-SZ survey. The ACTpol survey used the QE approach to detect the `halo' lensing of the CMB \citep{act_cmass2015} due to several thousands of massive ($M_{_{200}} \sim 2 \times 10^{13}\ M_{\odot}$) galaxies \citep{cmass_sample} detected from the Baryon Oscillation Spectroscopic Survey (BOSS) of the Sloan Digital Sky Survey (SDSS) III. The Planck Collaboration also reported CMB galaxy cluster lensing \citep{planckXXIV2015} due to 439 SZ selected clusters using a slightly modified QE approach \citep{melin2015}. 

The detection significance for CMB-cluster lensing depends on (i) the resolution of the telescope, (ii) the quality of the CMB data used (i.e.~noise level of the maps), and (iii) the number of clusters in the sample. Subsequently, the detections so far have focused on the brighter temperature lensing signal. However while fainter and being limited by statistical uncertainties currently,  polarization does have one significant advantage. On cluster scales (of order arc-minutes), the polarized astrophysical foregrounds are much fainter relative to the CMB than in temperature. The CMB polarization estimators are therefore more robust against astrophysical biases and effects. The astrophysical signals from galaxy clusters that may induce bias to the CMB temperature analyses include: the thermal Sunyaev-Zel{'}dovich (tSZ) \citep{SZ1970} effect which is $\mathcal{O}(10)$ higher than the lensing signal ($\sim 100 \ \mu K$); the kinematic Sunyaev-Zel'dovich (tSZ) \citep{kSZ1980} effect due to the motion of the cluster which is comparable to the lensing signal \citep{maturi2005, lewis2006, melin2015}; and the expected over-abundance of dusty star-forming galaxies (DSFGs) in the galaxy cluster \citep{shirokoff2010}.Other astrophysical signals that are uncorrelated with the cluster, including both uncorrelated power from the three sources above and radio galaxies,  contribute to an effective noise floor in the temperature maps, but should not bias the mass recovery. Except for the kSZ effect, these signals have a different spectral dependence than the CMB and may be removed in a linear combination of data from multiple frequencies. However, this cleaning will reduce the SNR, and if imperfect may bias the results.The level of foreground contamination is significantly lower for polarization maps which makes polarization extremely valuable for CMB-cluster lensing measurements from future low-noise CMB surveys.  

In this paper we construct a ML lensing estimator using a pixel-space likelihood approach \citep{baxter2015} from both temperature and polarization maps of the CMB. We then study and quantify potential systematic biases in the recovered cluster masses. Finally, we present forecasts for the performance of several upcoming CMB experiments like SPT-3G, AdvACT, Simons Array, and CMB-S4. The paper is organized as follows. In \S\ref{sec:method} we describe the MLE, pixel-pixel covariance matrix, and the cluster convergence profile used for lensing. The performance of the MLE is shown in \S\ref{sec_results}, along with a comparison to the performance of the QE. Next we examine various systematic errors in \S\ref{sec_sys_bias_checks}. Finally, we show forecasts for upcoming experiments in \S\ref{sec_forecast} before concluding in \S\ref{sec_conclusion}. 

\section{Method}
\label{sec:method}
In this section we explain the fundamentals of the temperature and polarization MLE for CMB-cluster lensing. We then describe the two major inputs to the MLE:   the calculation of the pixel-pixel covariance matrix, and  the cluster convergence profile used to lens the background CMB.

\subsection{Maximum likelihood estimators}
\label{sec_MLE_method}
The maximum likelihood approach works by fitting `lensed CMB' templates to the observed CMB maps. We follow \cite{baxter2015} in calculating the pixel-pixel correlations and performing the fitting in real space. We depart from past lensing polarization estimators  \citep{hu2007} by using Stokes Q/U maps instead of E- and B-mode maps. As we will discuss in \S\ref{sec_ideal_sims}, the two approaches yield nearly identical SNRs and Q/U maps have the advantage of being directly measured by experiments. To develop intuition in this work, we look separately at the temperature and polarization MLEs. Thus we calculate two pixel-pixel covariance matrices: one for the T maps and another for the combined Q/U maps. With this pixel-pixel covariance matrix $\Sigma_{lens}$ in hand, we can write down the  likelihood of obtaining the data as
\begin{eqnarray}
-2\ ln \mathcal{L}(\textbf{\textrm{d}}|\Sigma_{lens}) & = & ln\left| \Sigma_{lens} \right| + \textbf{\textrm{d}}^{T}\ \Sigma_{lens}^{-1}\ \textbf{\textrm{d}}
\label{eq_likelihood_single_cluster}
\end{eqnarray}
where the data vector $\textbf{\textrm{d}}$ is the pixel values of the observed T or Q/U maps. The pixel temperatures are defined as the variations from the mean CMB temperature and hence have zero mean.

We restrict our analysis to a $10' \times 10'$ box centered on the cluster to simplify and speed up the analysis. Since the majority of the lensing signal is near the cluster center,  this restriction has little impact on the lensing results. Doubling the box area to $14' \times 14'$ changes the SNR by $\le 1 \%$. Finally, in order to achieve a reasonable SNR, we need to combine the likelihoods for many clusters. In general, this can be done by taking a weighted average with weights $w_i$ of the cluster masses within a sample: 
\begin{eqnarray}
-2\ ln \mathcal{L}(\textbf{\textrm{d}}|\Sigma_{lens})_{tot} = & \sum\limits_{i = 0}^{n} w_{i}\left[ln\left| \Sigma_{lens} \right| + \textrm{\textbf{d}}_{i}^{T}\ \Sigma_{lens}^{-1}\ \textrm{\textbf{d}}_{i}\right]
\label{eq_combined_likelihoods}
\end{eqnarray} In this work, we use uniform weights as every simulated cluster has the same mass and redshift. 

\subsection{Pixel-pixel covariance matrix}
\label{sec_pixel_pixel_cov_matrix}

Besides the data, the only input to Eq.(\ref{eq_combined_likelihoods}) is the pixel-pixel covariance matrix $\Sigma_{lens}$ that encapsulates the lensing-induced mode-mode correlations. We calculate this pixel-pixel covariance matrix using a set of simulated skies:
\begin{eqnarray}
\Sigma_{lens}(M,z) & = & \left<(\textrm{\textbf{G}} - \left<\textrm{\textbf{G}}\right>) (\textrm{\textbf{G}} - \left<\textrm{\textbf{G}}\right>)^{T}\right>\nonumber\\
& = &  \frac{1}{n-1}\sum\limits_{i = 0}^{n} (\textrm{\textbf{G}}_{i} - \left<\textrm{\textbf{G}}\right>) (\textrm{\textbf{G}}_{i} - \left<\textrm{\textbf{G}}\right>)^{T},
\label{eq_lensed_cmb_cov_mass_z}
\end{eqnarray} where vector $\textrm{\textbf{G}}_{i}$ is the $10' \times 10'$ central box of the i$^{\rm th}$ sky realization. The required number of skies scales with the number of degrees of freedom in the covariance. In our case, the maximum number of degrees of freedom is for the Q/U estimator, for which the covariance matrix is an $800 \times 800$ matrix. We find that 130,000 realizations is adequate for recovering the cluster masses without bias. To remove any possible bias in $\Sigma_{lens}^{-1}$ due to the limited number of simulations, we apply the very small correction factor $(n_{sims} - n_{d} - 1)/n_{sims}$ where $n_{sims}$ = 130,000 and the length of the data vector $n_{d}$ = 400 (800) for T (Q/U) \citep{hartlap2006}.

The simulated sky realizations, $\textrm{\textbf{G}}_{i}$, are used both in this covariance calculation, and later to test for systematic biases and to estimate the mass uncertainties. Briefly, every realization contains a CMB sky that has been lensed by a galaxy cluster, and white instrumental noise.  For many uses, the realizations also include astrophysical signals from the kSZ effect, tSZ effect, DSFGs, and radio galaxies. Some of these astrophysical signals are uncorrelated with the galaxy cluster, such as radio galaxies or the tSZ effect from other haloes, while others are sourced in the cluster itself, such as the cluster's own tSZ signal. We provide more detail on the creation of these realizations in the Appendix \ref{sec_appendix}.

To evaluate the likelihood, this pixel-pixel covariance matrix needs to  be estimated as a function of mass and redshift. In this work, since we use a single cluster mass, $M_{_{200}} = 2 \times 10^{14}\ M_{\odot}$, and in most cases a single redshift, $z = 0.7$,  we calculate covariance matrices for a grid of masses with a mass resolution of $2.5\times 10^{12} M_{\odot}$ and selected redshifts. However, this brute-force fine gridding may not be computationally viable for real datasets where the cluster sample spans a wide range of mass and redshift. One proposed solution is to interpolate the covariance matrix from a coarse grid in $(M,z)$ space \citep{baxter2015}.

 \subsection{Cluster convergence profile}
 \label{sec_cluster_profile}
With the simulated CMB skies in place, the next step is to lens them using a convergence profile. Here we describe the numerical method to create the cluster convergence profile. Given a 2D lensing potential $\phi$, the convergence field can be obtained by calculating the divergence of its gradient $2\kappa(x) = -\nabla^{2}\phi(x)$. For an axially symmetric lens, the convergence at a radial distance $x$ from the cluster center can also be calculated as the ratio of the surface mass density of the cluster and the critical surface density of the universe at the cluster redshift $\kappa(x) = \frac{\Sigma(x)}{\Sigma(crit)}$. The critical surface density of the universe at cluster redshift is 
\begin{eqnarray}
\Sigma(crit) = \frac{c^{2}}{4\pi G}\ \frac{D_{_{\rm CMB}}}{D_{_{\rm clus}}D_{_{\rm CMB, clus}}}
\end{eqnarray} where $D_{_{\rm CMB}}, D_{_{\rm clus}}$ are the comoving distances to CMB ($z=1100$) and galaxy cluster ($z = 0.7$); and $D_{_{\rm CMB, clus}}$ is the distance between the CMB and the cluster. 
The surface mass density or the projected density of the halo is obtained by the line-of-sight integration of the halo density profile $\rho(r)$
\begin{eqnarray}
\Sigma(x) & = &   2 \int_0^{\infty} \rho(r)\ ds
\label{eq_surface_denisty_los_integral}
\end{eqnarray} where x is the distance to cluster center in the plane, r is the 3D distance to the cluster center and s is the distance along the line of sight with s=0 in the plane of the cluster. 

%$\delta_{c}$ is the dimensionless characteristic over-density. 
In this work, we define all quantities of the galaxy cluster with respect to the radius $R_{200}$ defined as the region within which the average mass density is 200 times the critical density of the universe $\rho_{crit}^z$ at the redshift of the halo.  Subsequently, the mass $M_{_{200}}$ within $R_{_{200}}$ will be
\begin{eqnarray}
M_{_{200}} = \frac{4 \pi}{3}\ R_{_{200}}^3\ \rho_{_{200}} = \frac{800 \pi}{3}\ R_{_{200}}^3\ \rho_{crit}^z
\label{eq_M_{_{200}}_1}
\end{eqnarray}
The mass $M_{_{200}}$ can also be found by directly integrating the expression for the density profile% in Eq.(\ref{eq_nfw_density})
\begin{eqnarray}
\label{eq_M_{_{200}}_2}
M_{_{200}} = \int\limits_0^{R_{_{200}}}4 \pi x^{2} \rho(x) dx = 4\pi  \int\limits_0^{R_{_{200}}} x^{2} \rho(x) dx
%= 4\pi \rho_{_{s}} \int\limits_0^{R_{_{200}}} \frac{r^{2}} {\left(\frac{r}{R_s}\right) \left(1 + \frac{r}{R_s}\right) } dr \nonumber\\
%& = & 4\pi \rho_{_{s}} R_s^3\int_0^{R_{_{200}}} \frac{r} {r + R_s} dr \nonumber\\
%& = & 4\pi \rho_{_{s}} R_s^3 \left[ ln(1+c) - \frac{c}{[1+c}\right]
%= 4\pi \rho_{_{s}} R_s^3 \left[ ln(1+c_{_{200}}) - \frac{c_{_{200}}}{[1+c_{_{200}}}\right]\nonumber
\end{eqnarray}

\noindent\textbf{NFW case:} We now apply this general framework for a spherically symmetric halo to  the specific case of a Navarro-Frenk-White (NFW) dark matter halo profile \citep{nfw1996} which has been used throughout this work to model the galaxy clusters unless specified otherwise. A NFW halo profile can be characterized by its scale radius $R_{s}$, the dimensionless concentration parameter $c$, and the dimensionless characteristic over-density $\delta_{c}$. The characteristic over-density is defined by $\delta_c = \rho_0 /\rho_{crit}^z$, where $\rho_0$ is the central cluster density and $\rho_{crit}^z$ is the critical density of the Universe at redshift z. In terms of these quantities the NFW halo density profile is written as
\begin{eqnarray} 
\rho\left(r\right) & = & \frac{\delta_{c}\rho_{crit}^z}{\left(\frac{r}{R_s}\right)\ \left(1+\frac{r}{R_s}\right)^2}.
\label{eq_nfw_density_with_delta_c}
\end{eqnarray} 
The goal now is to insert this expression into Eq.(\ref{eq_surface_denisty_los_integral}) to obtain the convergence field of the NFW halo. 
We can change the variables using $s = \sqrt{r^2 - x^2 }$ and $ds = \frac{rdr}{\sqrt{r^2 - x^2}}$ to yield:
\begin{eqnarray}
\Sigma(x) =  2 \delta_c \rho_{crit}^z R_s^3 \int_x^{\infty} \frac{1}{r (R_s + r)^2} \frac{r\ dr}{\sqrt{r^2 - x^2}}
\label{eq_surface_density_nfw}
\end{eqnarray} The only quantity that remains to be defined is the scale radius $R_{s}$ which we get by 
\begin{eqnarray}
c \equiv c_{_{200}} = \frac{R_{_{200}}}{R_s}
\label{eq_conc_parameter}
\end{eqnarray} In this work we set $c_{_{200}}=3.0$ following \cite{Bhattacharya2013}. 
The solution to Eq.(\ref{eq_surface_density_nfw}) can also be solved analytically and an explicit closed-form expression  for the NFW case has been given by \cite{bartelmann1996}. The general framework described here allows one to obtain the convergence profile for other halo density profiles. In Section~\ref{subsec_systematics_cluster_profile} we explore lensing using other plausible halo density profiles.

%%%%%%%%%%%%%%%%%%%%%%%%%%%%%%%%%%%%%%%%%%%%%%%%%%%%%%
%%%%%%%%%%%%%%%%%%%%%%%%%%%%%%%%%%%%%%%%%%%%%%%%%%%%%%
%%%%%%%%%%%%%%%%%%%%%%%%%%%%%%%%%%%%%%%%%%%%%%%%%%%%%%

\section{Results}
\label{sec_results}
In this section we report the expected uncertainties on the recovered cluster masses from the temperature and polarization estimators 
for an idealized no astrophysical foregrounds scenario, and compare the results for the MLE to those for the QE \citep{hu2007}. We then study the extent to which foregrounds degrade the lensing mass uncertainties. 
%before and the effects due to cleaning the frequency dependent foregrounds \emph{via} internal linear combination methods. 
%after the addition of foregrounds. 
%We then compare the MLE to the standard QE \citep{hu2007}.% Finally, we discuss the systematic bias in the recovered lensing mass expected from a comprehensive list of sources.
Throughout this work, we quote mass uncertainties on a fiducial cluster sample containing 100,000 clusters at redshift $z=0.7$ and mass \mbox{$\mvar{} = 2 \times 10^{14}\ M_{\odot}$}. This sample size is very approximately the number expected for the CMB-S4 experiment.
We calculate the mass uncertainties $\Delta \mvar{}$ by looking at the log-likelihood ($-2\ ln \mathcal{L}$) as a function of mass. According to Wilks' theorem, the statistic $-2\ ln \mathcal{L}$ follows a $\chi^2$ distribution with one degree of freedom (corresponding to one free parameter $M_{_{200}}$). 
Thus, we calculate the mass uncertainty by finding a change of $\Delta \chi^2 = 1$ relative to the best-fit point:
\begin{eqnarray*}
\Delta \chi^2 = \chi^2_{_{M_{est}}} - \chi^2_{_{\mvar{}}} = -2\ ln \mathcal{L}(M_{est}) + 2\ ln \mathcal{L}(\mvar{}).
\end{eqnarray*}
We report the mass uncertainties, $\Delta \mvar$, as a function of temperature map noise levels in $\ukarcmin$ with the corresponding Q and U polarization map noise levels assumed to be higher by $\sqrt{2}$. 
Note that foreground cleaning will generally break this simple relationship. 
As the dominant foregrounds are largely unpolarized, the foreground-cleaned temperature map  noise levels  will be relatively higher.

\begin{figure}
\centering
\includegraphics[width=0.8\textwidth, height=0.75\textwidth]{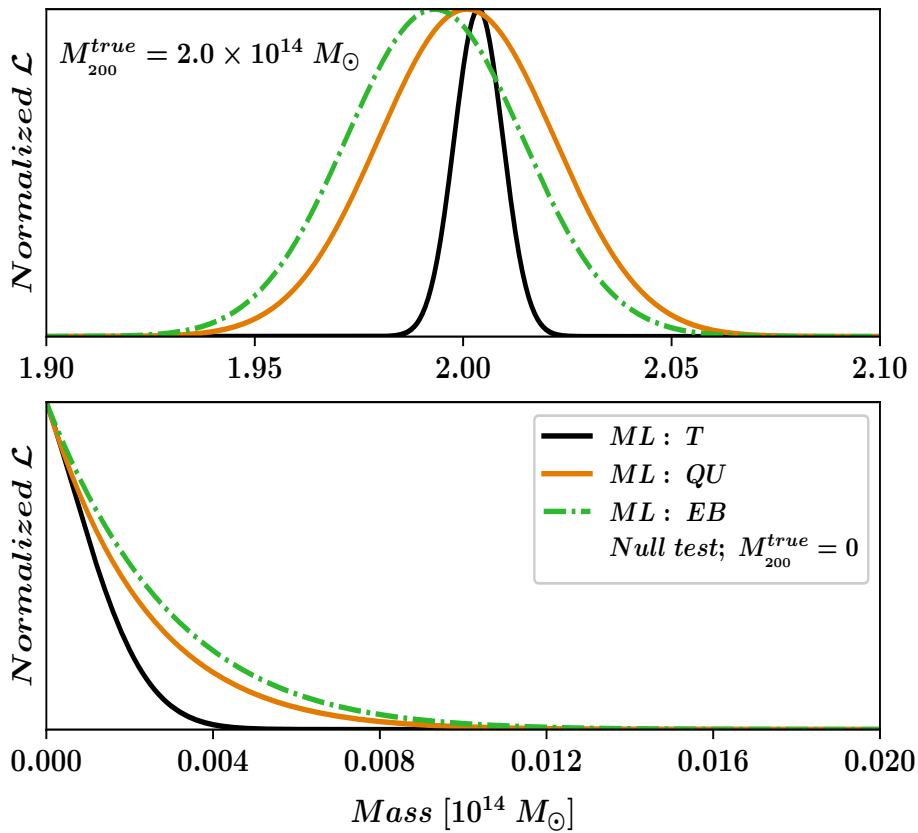}
\caption{ The MLEs recover the true cluster masses. 
The top panel shows the normalized likelihood curves for the $T_{\rm ML}$ (black solid), $QU_{\rm ML}$ (orange solid), and $EB_{\rm ML}$ (green dash dotted) MLE for a cluster sample containing 100,000 clusters with mass $M_{_{200}} = 2 \times 10^{14}\ M_{\odot}$ (red solid line) at redshift $z=0.7$. 
The performance of the $QU_{\rm ML}$  and $EB_{\rm ML}$ estimators  is comparable. 
The bottom panel shows the results for a null test for fields with no galaxy clusters; as expected the likelihood peaks at zero mass. 
All curves are for an idealized case with no astrophysical foregrounds, a temperature map noise level of $1\ \ukarcmin$, and a beam size of $\theta_{\rm FWHM} =1^\prime$.
 }
\label{fig_likelihood_1000_clusters_100_sims_T_QU_EB_comp}
\vspace*{2mm}
\end{figure}

\subsection{Ideal simulations: Performance comparisons}
\label{sec_ideal_sims}

We begin by looking at an idealized case: CMB-only simulations with no Galactic or extragalactic foregrounds. 
This idealized case serves two purposes. 
It provides a benchmark to understand the effects of these foregrounds, and it allows the performance of the MLE to be compared under equivalent assumptions to past results for the standard QE \citep{hu2007}. % and iterative QE \citep{yoo2008, yoo2010}. 
In Fig.~\ref{fig_likelihood_1000_clusters_100_sims_T_QU_EB_comp} we show the combined likelihood curves of 100,000 clusters (\emph{cf.} Eq.(\ref{eq_combined_likelihoods})) for the temperature $T_{\rm ML}$ (black solid) and polarization $QU_{\rm ML}$ (orange solid), $EB_{\rm ML}$ (green dash dot) MLEs. The top panel shows that all the estimators recover the true cluster mass (red solid line). 
At a noise level of $\Delta T = 1$\,\ukarcmin{}, the no-lensing case is rejected at a significance ($\sqrt{\Delta \chi^{2}}$) of 400\,$\sigma$ for the temperature MLE and 110\,$\sigma$ for the polarization MLE. 
 The bottom panel of Fig.~\ref{fig_likelihood_1000_clusters_100_sims_T_QU_EB_comp} shows a null test where the estimators are applied to fields without galaxy clusters. The likelihoods peak at zero mass consistent with no lensing. 

\begin{figure}
\centering
\includegraphics[width=1.\textwidth, height=0.75\textwidth]{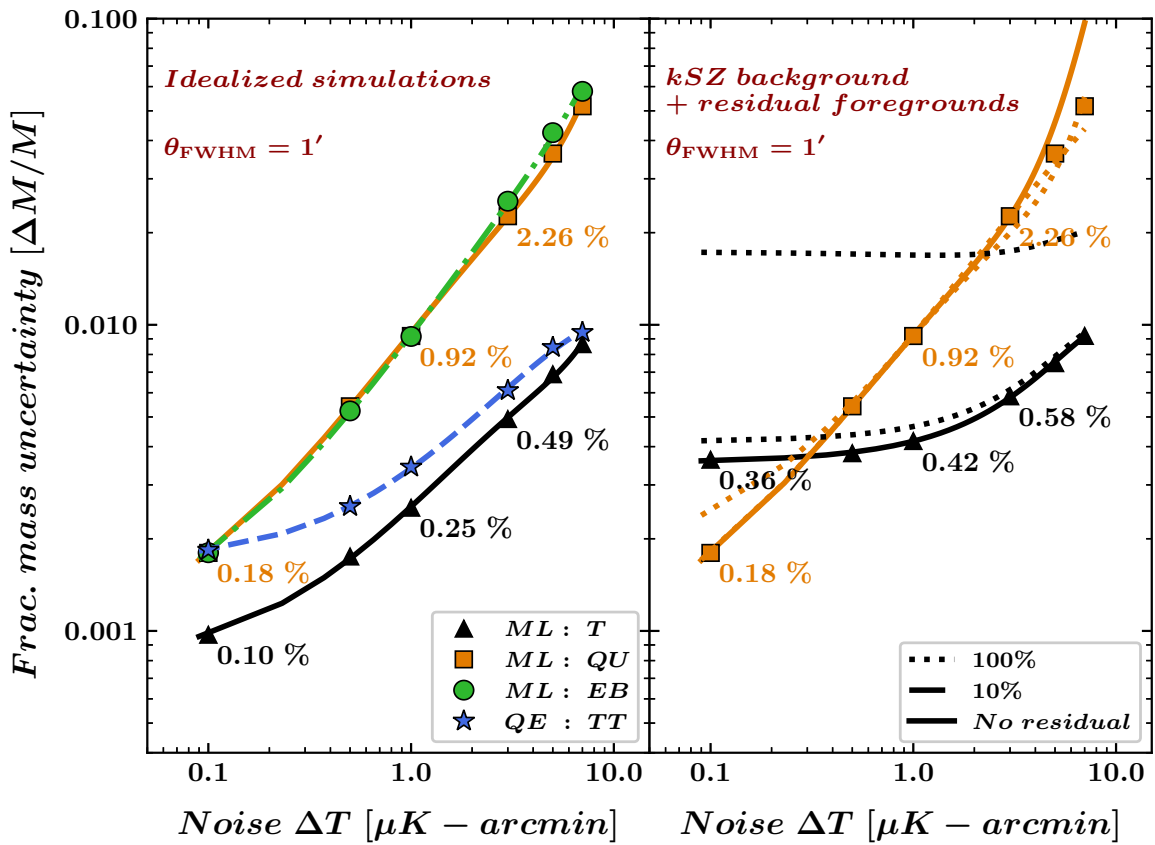}
%\caption{Fractional cluster mass uncertainties with a sample containing 100,000 clusters. {\bf \emph{Left panel:}} In an idealized CMB-only case,  temperature is the main information channel for CMB-cluster lensing. We caution the reader that this conclusion changes when foregrounds and systematics are considered. In this case, $T_{\rm ML}$ (black triangles) out-performs a standard QE (blue stars) at the map noise levels expected for upcoming CMB experiments. As expected, polarization estimators $QU_{\rm ML}$  (orange squares) and $EB_{\rm ML}$  (green circles) perform nearly identically. An experimental beam of $\theta_{_{\rm FWHM}} = 1'$ has been assumed here. {\bf \emph{Right panel:}} Extragalactic foregrounds can have a large impact on the mass uncertainties from $T_{\rm ML}$, but do not significantly affect $QU_{\rm ML}$. The lines correspond to different level of residual foreground contamination: solid (0\%), dash-dot (10\%), and dotted (100\%). An uniform kSZ background \mbox{$D_{\ell=3000}^{^{\rm kSZ}} = 2.9\, \mu K^{2}$} \citep{george2015}, which cannot be cleaned by combining data from different frequency channels, has been assumed for all the curves. An experimental beam of $\theta_{_{\rm FWHM}} = 2'$ has been used here.}
\caption{Fractional cluster mass uncertainties with a sample containing 100,000 clusters. {\bf \emph{Left panel:}} In an idealized CMB-only case,  temperature is the main information channel for CMB-cluster lensing. We caution the reader that this conclusion changes when foregrounds and systematics are considered. In this case, $T_{\rm ML}$ (black triangles) out-performs a standard QE (blue stars) at the map noise levels expected for upcoming CMB experiments. As expected, polarization estimators $QU_{\rm ML}$  (orange squares) and $EB_{\rm ML}$  (green circles) perform nearly identically. {\bf \emph{Right panel:}} Extragalactic foregrounds can have a large impact on the mass uncertainties from $T_{\rm ML}$, but do not significantly affect $QU_{\rm ML}$. The lines correspond to different level of residual foreground contamination: solid (0\%), dash-dot (10\%), and dotted (100\%). An uniform kSZ background \mbox{$D_{\ell=3000}^{^{\rm kSZ}} = 2.9\, \mu K^{2}$} \citep{george2015}, which cannot be cleaned by combining data from different frequency channels, has been assumed for all the curves. An experimental beam of $\theta_{_{\rm FWHM}} = 1'$ has been used in both cases.}
\label{fig_delM_M_1000_clusters_T_QU_EB_ideal_FG}
\vspace*{2mm}
\end{figure}

In left panel of Fig.~\ref{fig_delM_M_1000_clusters_T_QU_EB_ideal_FG} we show the lensing mass uncertainties for the temperature and polarization estimators. One takeaway from this figure is that without foregrounds, temperature (black triangles) would be the primary channel for CMB-cluster lensing from a raw SNR perspective. 
The polarization $QU_{\rm ML}$ (orange squares) estimator only dominates below $0.075\,\ukarcmin$, a noise level which is unlikely to be achieved even with the proposed CMB-S4 experiment. 
The relative performance of the temperature and polarization estimators can be understood as (1) the lensing signal scales with the amplitude of the background gradient which is $\sim$10x brighter in temperature, but (2) as the noise levels drop the background CMB acts as an additional noise source for the temperature estimator which limits its performance. 
We remind the reader however that this is an idealized case, and, as will be shown later, the polarization channels have advantages with respect to systematic biases and astrophysical foregrounds. 

This figure also shows that MLEs using Q/U maps (orange squares) and E/B-mode maps (green circles) achieve comparable SNRs. 
There is little to differentiate between the two approaches from a theoretical standpoint. The apparent difference between the Q/U and E/B estimators at high map noise levels is not statistically significant. 
However, working with the  $QU_{\rm ML}$ estimator simplifies the analysis by eliminating the coordinate transformation from Stokes Q/U to E/B-mode maps. 
We choose to proceed with the $QU_{\rm ML}$ polarization estimator in this work. 

Finally,  we compare the performance of the temperature MLE (ML: T,  black triangles) and QE (QE: TT, blue stars). 
The standard QE is a linear approximation to the MLE and performs similarly at low SNR values.
While not shown,  the polarization MLE and QE perform equivalently for the range of map noise levels plotted as they remain in the low SNR regime. 
As the map noise levels decrease, the temperature MLE begins to out-perform the QE. 
The extra information  exploited by the MLE can also be recovered using an iterative version of the QE as demonstrated by  \citet{yoo2008, yoo2010}. 
We find a factor of $\sim2$ improvement in the MLE uncertainties compared to the QE at 0.1\,\ukarcmin{} for our fiducial sample of $\mvar = 2 \times 10^{14}\ M_{\odot}$ and redshift $z = 0.7$. 
Using a MLE or iterative QE dramatically improves the performance of the temperature estimators at low map noise levels.

A difficulty for the QE and MLE for cluster masses is that both estimators use an assumed cluster mass profile. 
This dependence shows up in different places in each analysis:
\begin{itemize}
\item The QE extracts the convergence field by exploiting the lensing-induced correlation between the gradient of the temperature/polarization field and the unlensed background. 
This step is model-independent. 
However, to improve the SNR, cluster profiles are then fit to the convergence field in order to estimate the galaxy cluster masses. 
\item In the MLE, the mass is obtained by directly fitting lensing templates of known cluster profiles at the map level using the full pixel-space likelihood. 
\end{itemize}
As both methods suffer from the same model dependence, there is no reason not to take advantage of the MLE's improved performance at low noise levels.

%\clearpage

\subsection{Effects of  astrophysical foregrounds}
\label{sec_foregrounds}

%Extragalactic and Galactic foregrounds will introduce additional noise in, and for signals originating in the cluster itself, likely bias the lensing estimators. 
Extragalactic and Galactic foregrounds will introduce additional noise in the lensing estimators, and, for signals originating in the cluster itself, they will likely also bias the lensing estimators.
In this work, we neglect Galactic foregrounds under the assumption that an appropriate frequency combination can effectively suppress the Galactic contamination. 
 Since the level of the foregrounds is much higher for temperature than polarization, the foreground `noise' will primarily affect $T_{\rm ML}$. 
 Here we look at the impact of foregrounds on 150\,GHz maps, considering the tSZ effect, kSZ effect, radio galaxies, and dusty galaxies. 
 Our model for these signals is described in the Appendix \ref{sec_appendix_extragal}. 
 We present the effect of these signals on the mass uncertainties here, while the related systematic biases are shown in \S\ref{sec_sys_bias_checks}.

The effect of extragalactic foregrounds on the mass uncertainties is displayed in the right panel Fig.~\ref{fig_delM_M_1000_clusters_T_QU_EB_ideal_FG} for different levels of foreground cleaning. 
While essentially negligible for the polarization estimators, astrophysical foregrounds are very important for the temperature channel. 
Without any cleaning (dotted lines), foregrounds set an effective noise floor around $\sim 5\,\ukarcmin$, with the mass uncertainty from $T_{\rm ML}$ (black triangles) thus plateauing at 1.8\%. 
The polarization channel, $QU_{\rm ML}$ (orange squares), out-performs the temperature channel at noise levels below 1.5\,\ukarcmin{} in this case. 
We can of course, combine multiple frequency channels to suppress these foregrounds at the cost of increasing the noise levels in the temperature map. 
The exact effects depend on the relative noise levels and beam sizes in each frequency band for a specific experimental configuration. 
While we take this into account properly in  \S\ref{subsec:cmbs3s4} when forecasting the performance of planned CMB experiments, here we build intuition into the required foreground cleaning with a simplified model. 
Namely, we use a Gaussian FWHM = $1'$ beam and assume no changes to the it due to foreground cleaning, perfect suppression of the tSZ power, and no suppression of the kSZ power. 
Furthermore, we assume that the foreground cleaning leaves behind 100\% (dashed lines), 10\% (dot-dashed lines) or 0\% (solid lines) of the radio and dusty galaxy power. 
 With 90\% foreground cleaning, temperature once again becomes very competitive at the expected noise levels of CMB experiments in the near-term -- even after allowing for a factor of a few increase in map noise levels due to cleaning. 
 Perfect foreground cleaning only slightly improves the SNR beyond the 90\% case. 
Note that foreground cleaning is also likely to enlarge the effective experimental beam which will further degrade the mass uncertainties, an effect which is not captured in this plot. 

In summary, extragalactic foregrounds have little impact on the polarization estimators, but significantly increase the mass uncertainties in the temperature channel if not cleaned. % in some fashion. 
Additionally, as will be shown in the next section, these astrophysical signals lead to some challenging systematic biases for the  temperature estimator. 

\section{Systematic bias checks}
\label{sec_sys_bias_checks}
We now turn from the statistical power of the cluster lensing MLE to the magnitude of potential systematic biases. 
We examine the following sources of systematic error:
\begin{enumerate}
\item[1.] Chance superpositions of other halos near the cluster position, 
\item[2.] Differences between the assumed and true cluster mass profiles,
\item[3.] Uncertainty in the cluster position, 
\item[4.] Uncertainty in the cluster redshift. 
\item[5.] The kSZ signal from the cluster (temperature-only), 
\item[6.] The tSZ signal from the cluster (temperature-only), and 
\item[7.] Dusty galaxies residing in the galaxy cluster.% (temperature-only).
\end{enumerate} 
The last three effects are largely unpolarized, and have little impact on polarization lensing estimators. 
Note that even sources of bias that have identical effects on polarization and temperature can have a different bias in the two estimators due to  differences in mode weighting between  estimators. 

We report on the bias in an idealized survey (unless specified otherwise) with $1^\prime$ beam and temperature noise levels of $1\,\ukarcmin$ (and polarization noise higher by $\sqrt{2}$). 
No foreground terms are included unless noted. 
The magnitude of the bias from each source does depend on these assumptions as the beam size, map noise, and foreground levels  determine the weighting of different angular scales in the MLE. 
The effect of systematic error is simulated on a  sample of 1,000,000  clusters with a mass $M_{_{200}}^{true} = 2 \times 10^{14}\ M_{\odot}$ and, unless noted otherwise, a redshift of 0.7.
 The bias is calculated from this sample according to:
\begin{eqnarray}
\label{eq_sys_bias_calc}
b = \frac{M_{200}^{sys}}{M_{{200}}^{true}} -1,
\end{eqnarray} 
where  $M_{{200}}^{sys}$ is the average recovered lensing mass. 
We also estimate the error on the bias by looking at its scatter across 100 subsamples.

The resulting bias estimates for the seven sources are summarized in Table \ref{tab_sys_bias} and discussed in detail below. With the exception of the redshift uncertainties, a careful accounting of the bias from all of these sources will be necessary to achieve the percent-level mass calibration hoped for from CMB-S4. Other sample-specific systematic errors may also be a concern, for instance, the number of false positive in the catalog, selection effects, or some form of bias in the weighting of clusters when combining likelihoods in Eq.\ref{eq_combined_likelihoods}. 

\subsection{Chance superpositions with other haloes}

CMB lensing is sensitive to the mass integrated along the line of sight, which means the lensing-derived mass of a given cluster will include contributions from any other haloes near the line of sight and correlated to the cluster, normally described as the {\it 2-halo term} \citep{seljak2000,cooray2002}. This is a particular concern for lower-mass haloes which may not be detected by the cluster survey. 
To check the effects of such low mass haloes, we draw upon the halo catalog for the publicly available N-body simulations of \citet{flender2016}. 
This catalog contains $\sim 17$ million objects; we begin by selecting the $\sim$\,20,000 haloes in the catalog that match the baseline cluster ($\mvar \in [1.8,2.2]\times10^{14}\ M_{\odot}; z \in[0.6,0.8]$). 
When creating each mock map, we select one of these 20,000 haloes, and add convergence profiles based on the mass and redshift of any other nearby haloes (within 50$^\prime$) to the central cluster's convergence profile. 
We lens the CMB by these convergence profiles. 
From there, the analysis proceeds as normal: the lensed CMB maps are convolved with the beam and noise is added.
%Nearby, lower-mass  haloes bias the recovered mass high by a few percent: $2.5 \pm 0.3$\% and $6.3 \pm 1.1$\% for $T_{\rm ML}$ and $QU_{\rm ML}$ respectively. Accurate and reliable simulations of structure formation on large volumes will be crucial to handling this source of systematic error at the sub-percent level.
Nearby, lower-mass haloes bias the recovered mass high by a few percent:  $2.5 \pm 0.3 \%$ and $6.3 \pm 1.1 \%$ for $T_{ML}$ and $QU_{ML}$ respectively. This bias can be mitigated by explicitly fitting for the {\it 2-halo term} \citep{johnston2007, oguri2011}. This, however, will slightly worsen the constraints on the recovered lensing mass because of the additional nuisance parameters involved in the fitting process.
In addition, there could also be other plausible biases that can originate since clusters are not isolated structures but preferentially located on large-scale filaments. For example, clusters with ellipticity aligned along the line of sight, will have higher tSZ surface brightness and therefore more likely to be selected. But they are also more likely to be part of a filament that is along the line of sight, and therefore their inferred lensing mass would tend to be overestimated. A detailed study of this is beyond the scope of this paper but we would like to caution the reader as these effects can be challenging when one is aiming for a $\sim 1\%$ level mass precision with CMB-S4. Clearly, accurate and reliable simulations of structure formation on large volumes will be crucial to handling this source of systematic error at the sub-percent level.

\subsection{Cluster mass profiles}
\label{subsec_systematics_cluster_profile}
As previously mentioned, both the QE and MLE use an assumed cluster mass profile, and any difference between the assumed profile and true average profile will bias the inferred masses. In this work, we have used the NFW profile. However, the NFW profile is known to be an imperfect approximation and previous studies of very massive clusters have observed significant deviations from an NFW profile at $r \gtrsim 0.5R_{{200}}$ \citep{diemernfw_deviation}. It is unknown whether these deviations will be larger or smaller for the lower cluster mass assumed  in this work. For reference, the angular size of the clusters used in this work is $\theta_{{200}} = D_{L} R_{{200}} = 2.^\prime2$. %and the assumed beam size throughout this section is $\theta_{\rm FWHM} = 1^\prime$
To estimate the  size of this bias, we consider three alternative cluster profiles: 
\begin{enumerate}

\item A modified version of the NFW profile which drops off more rapidly with radius
\begin{eqnarray}
\kappa_{\rm NFW}^{mod}(x) =
\begin{cases}
    \kappa_{\rm NFW} &; x \le 0.75\theta_{_{200}}\\
    \kappa_{\rm NFW} \times m(i,j)&; 0.75\theta_{_{200}} < x \le 1.5\theta_{_{200}}\\
    0 &; otherwise
\label{eq_kappa_mod}
\end{cases}
\end{eqnarray} where $m(i,j)$ is a Hanning 2D apodization kernel. 
We create the 2D apodization kernel as $m(i,j) = m(i) \times m(j)$ with $m(i) = \frac{1}{2} \left[ 1 - cos\left(\frac{2\pi  (i-n/2) }{n} \right)\right]$, where i and j are pixel indices in the $n \times n$ map.

\item A change to the cuspiness of cluster core in the NFW profile, following \citet{king2001}
\begin{eqnarray}
\kappa_{\rm NFW}^{sub}(x)  & =  &
\begin{cases}
    \kappa_{\rm NFW}\ + \sum\limits_{i=1}^{3} \kappa_{sub}^{i}&; x \le 1'\\
    \kappa_{\rm NFW} &; otherwise;
\label{eq_kappa_sub}
\end{cases}
\end{eqnarray}

\item The Einasto profile \citep{einasto1989}
\begin{eqnarray}
\rho(r)_{_{Ein}} & = &  \rho_{_{0}}\ exp\left( - \frac{2}{\alpha} \left[\left(\frac{r}{R_s} \right)^{\alpha} - 1\right]\right)
\label{eq_einasto_density}
\end{eqnarray}
with the shape parameter $\alpha = 0.18$ \citep{ludlow2013}. The convergence profile for this profile is obtained by inserting this into into Eq.(\ref{eq_surface_denisty_los_integral}).

\end{enumerate}
We simulate lensing by each of these three convergence profiles, and then look at the recovered mass when incorrectly assuming the NFW profile. The wrong cluster mass profile biases the recovered cluster masses from -2.5\% to 3.8\% as shown in Table \ref{tab_sys_bias}. 
A similar  test has been performed by \cite{sereno2015} for optical weak-lensing analysis. 
They find biases from -2\% to 7\% depending on the cluster mass when an NFW profile is assumed instead of a true Einasto profile for the galaxy cluster, comparable to what is found here for CMB-cluster lensing. 
This level of bias is larger than the statistical mass uncertainties expected for the CMB-S4 experiment, and a significant challenge for upcoming experiments. 
More work will be required to accurately measure the cluster mass profiles as a function of radius if we are to achieve the full potential of galaxy cluster cosmology.

\begin{figure}
\centering
%\hspace{-2mm}
\includegraphics[width=0.85\textwidth, height=0.65\textwidth,clip=]{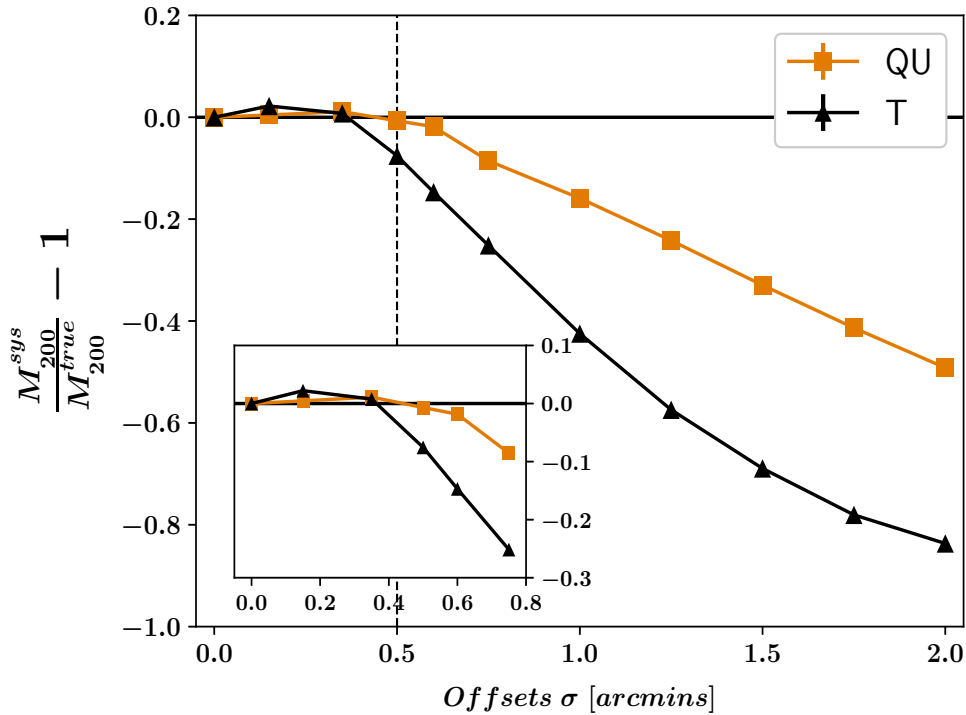}
\caption{The low mass bias due to the positional offsets between SZ determined centroid and the true cluster center  for the $T_{\rm ML}$ (black triangles) and $QU_{\rm ML}$ (orange squares) estimators. The typical positional offset of $0.5'$ \citep{song2012} is marked with the vertical dashed line. The error bars are derived using estimates gathered by repeating the test 100 times.}
\label{fig_sys_bias_cluster_offsets}
\vspace*{2mm}
\end{figure}

\subsection{Cluster positions}
If  the wrong position is used for a cluster (due to normal positional uncertainties), the lensing estimator will `miss' some mass and the estimated mass will be biased low. Here we quantify the magnitude of the bias as a function of the positional uncertainty. 
For each cluster in the sample, we draw an offset between the true and nominal cluster positions from a normal distribution, $N(0,\sigma^2)$.
The results as a function of the rms offset, $\sigma$, are shown in Fig.~\ref{fig_sys_bias_cluster_offsets}. 
The temperature MLE's bias is larger for a given $\sigma$ because the temperature MLE draws more information from small angular scales. 
This also implies that surveys with larger beams, higher noise levels or more residual foreground power would incur a smaller bias due to positional uncertainties.

An obvious question for interpreting Fig.~\ref{fig_sys_bias_cluster_offsets} is what a reasonable positional uncertainty would be for the upcoming experiments and detailed simulations are needed to accurately determine them.
%This question has been addressed for SZ-selected cluster catalogs, by comparing the SZ centroid to the X-ray centroid \citep{linden2014} or location of the brightest central galaxy (BCG)  \citep{song2012, saro2015}. In both cases, the typical offset is of order $0.5^\prime$, which is  consistent with the expected positional uncertainty due to the beam size and finite SNR of the SZ survey. If uncorrected, a positional uncertainty of $0.5^\prime$ would cause lensing masses for future CMB surveys to be significantly underestimated -- low by $7.5 \pm 0.3\%$ for temperature and $2.5 \pm 1.3\%$ for polarization. 
For a typical offset of $0.5^\prime$ between the SZ centroid and X-ray centroid \citep{linden2014} or location of the brightest central galaxy (BCG) \citep{song2012, saro2015}, the bias on the recovered lensing masses for future CMB surveys can be significantly underestimated: $7.5 \pm 0.3\%$ for temperature and $1.5 \pm 1.3\%$ for polarization. 
However, given external constraints on the expected positional uncertainty, a correction could easily be applied for this bias with a modest decrease in the SNR. 
Neglecting foreground power, one would need to know $\sigma$ to $\sim$\,2\% (8\%) to recover the true mass to within 1\% for the temperature (polarization) MLE. 
Adding residual foregrounds will tend to de-weight the small-scale temperature information and relax this requirement on how well the positional uncertainty must be known. 

\subsection{Redshift uncertainty}
The catalogs of clusters to be weighed are likely to come from SZ surveys (e.g., SPT-3G \citep{benson2015_3g}, AdvACTpol \citep{advact_2016}, CMB-S4 \citep{cmbs4_2016}), optical imaging surveys (e.g., DES \citep{rykoff2016}, LSST \citep{lsst_science_book}) or X-ray surveys (e.g., eRosita \citep{erosita_science_book}). 
The majority of clusters will not have spectroscopic redshifts given the prohibitive amount of observing time  required to do spectroscopy on tens or hundreds of thousands of clusters. Instead, we can expect to have red sequence redshifts up to an upper redshift threshold, and a redshift lower limit for clusters at even higher redshifts. The very high redshift case is particularly relevant  to the SZ surveys as the SZ surface brightness does not fall off with redshift. The current state of the art for red sequence redshifts can be seen in the  DES \texttt{redMaPPer} catalog, where the \emph{photo-z} errors are $\sigma_{z} = 0.01 (1+z)$ for $z \le 0.7$ and $\sigma_{z} = 0.02 (1+z)$ for $z \sim 0.9$ \citep{rykoff2016}. The mass bias is estimated by considering a redshift scatter for individual clusters. We conservatively take the redshift errors to be
\begin{equation}\nonumber
\sigma_z = \left\{ 
\begin{array}{l}
    0.02 (1+z);  ~~z < 1 \\
    0.06 (1+z);  ~~z > 1
  \end{array}\right.
\end{equation}

We create mock lensed CMB maps using the true redshifts $z\in[0.5, 1, 1.5, 2]$ for the cluster. However, the `measured' redshift, $z + \sigma_{z}$, for each cluster is used to construct the pixel-pixel covariance matrix. The masses are then fit using this incorrect covariance matrix, and the fractional bias determined. 
%The case that should have the largest bias, with the second model at $z=2$ is listed in Table~\ref{tab_sys_bias}. 
%We do not detect a mass bias at any redshift in either case.
The case that should have the largest bias, $z=2$, is listed in Table~\ref{tab_sys_bias}. We do not detect a mass bias at any redshift.

\subsection{Kinematic SZ signal}
\label{sec_kSZ_bias}
The motion of the galaxy cluster with respect to the CMB reference frame induces a Doppler shift in the CMB photons, an effect known as the kSZ effect. Unlike the tSZ effect, the kSZ effect has the same spectrum as the CMB and can not be removed using multiple frequency bands. 
Thus while the magnitude of the kSZ signal in a cluster is much lower than the tSZ signal at 150 GHz, the bias due to the kSZ is more intractable. The kSZ signal is effectively unpolarized, as argued in the Appendix \ref{sec_appendix_SZ}.

We estimate the bias due to the kSZ effect using the publicly available kSZ maps\footnote{\url{http://www.hep.anl.gov/cosmology/ksz.html}} and halo catalog from \citet{flender2016} simulations. These are full sky simulations in \texttt{HEALPix} pixelization scheme \citep{healpix} with nside = 8192 corresponding to pixel resolution of $0.42^\prime$. We begin by selecting the $\sim$\,20,000 haloes in the catalog that match the baseline cluster ($\mvar \in [1.8,2.2]\times10^{14}\ M_{\odot}; z \in[0.6,0.8]$). We extract $50^\prime \times 50^\prime$ regions around each halo and project them into the flat-sky approximation with a pixel size of $0.5^\prime$. The maps are then smoothed by a FWHM=1$^\prime$ Gaussian beam. One of these randomly chosen smoothed kSZ maps is then added to each mock dataset before fitting for the cluster masses. 
If we completely neglect the extra signal due to the cluster's own kSZ effect from the pixel-pixel covariance, the kSZ effect dramatically biases the temperature MLE: the inferred masses are low by 41\%. We get similar large bias levels when we used an analytic modelling for the cluster kSZ signal instead of extracting them from N-body simulations. Modelling the optical depth of the clusters using the Battaglia \cite{nickb2016} profile and drawing the cluster velocities from a normal distribution $N(0, \sigma^{2})$ with scatter $\sigma = 350\ km/s$, we obtained a 32\% bias in the recovered lensing mass.

Of course, one would presumably include the kSZ signal into the pixel-pixel covariance matrix. Indeed, the bias disappears as expected with perfect knowledge of the kSZ signal. %We demonstrate this by constructing the covariance matrix from the same set of \citet{flender2016} simulations as using in the mock data, finding no bias: b = $0.0 \pm 0.4\%$. 
We demonstrate this by finding no bias (b = $0.0 \pm 0.4\%$) when we construct the covariance matrix from the same set of \citet{flender2016} simulations used for the mock data creation. However, there is uncertainty of order 20\% in current predictions for the kSZ effect in a single cluster due to questions about the detailed physics of the intracluster medium. When we repeat the test with a 20\% residual kSZ, the recovered mass is biased low by 7.5 $\pm$ 0.4\%. The bias can either be high or low depending on whether we over- or under-estimate the residual kSZ signal when constructing the covariance matrix. Linearly interpolating between these points suggests that residual kSZ at the 2-3\% level would lead to percent level biases in the reconstructed mass. Given the current uncertainties in predicting the kSZ signal from a cluster of a given mass, the kSZ bias presents a serious obstacle for using temperature lensing information from future CMB experiments.

\subsection{Thermal SZ (tSZ) signal and tSZ cleaning}
\label{subsec:tszbias}

The tSZ effect is produced by inverse Compton scattering between the hot electrons of the intracluster medium and CMB photons. 
The tSZ signal is $\mathcal{O}(10)$ times higher than the expected lensing signal from a cluster, and  strongly biases the temperature lensing estimator if ignored. 
The polarized tSZ signal is negligible as argued in the Appendix \ref{sec_appendix_SZ}. 
We consider two approaches to handle the potential tSZ bias, (1) using multiple frequencies to separate the CMB and tSZ signals, and (2) including the tSZ signal in the pixel-pixel covariance matrix. 
Both approaches completely eliminate the tSZ bias in an ideal world, however as we will discuss,  imperfect knowledge of the experimental calibration or true tSZ signal can lead to a remaining bias.
Consistency between these approaches is a potential cross-check on any remaining tSZ bias.

To evaluate the performance of each approach,  we use Compton $y$ maps produced on a $5^{\circ} \times 5^{\circ}$ box at resolution $2.'5$ from the smoothed-particle hydrodynamics (SPH) simulations of  \citet{mccarthy2013}. 
Neglecting relativistic corrections, we convert the Compton $y$ maps into tSZ maps by
\begin{eqnarray}
\Delta T = y T_{\rm CMB} \left[x \left(\frac{e^{x/2} + e^{-x/2}}{e^{x/2} - e^{-x/2}}  \right)\right]
\end{eqnarray} where $x = \frac{h\nu}{k_{\rm B}T_{\rm CMB}}$, $\nu$ is the frequency in GHz, $T_{\rm CMB} = 2.73\ K$, $k$ is the Boltzmann constant, and $h$ is the Planck constant. 
We assume the experiment has two frequency bands centered at 90 and 150 GHz. 

\subsubsection{tSZ frequency cleaning} 
We first create a tSZ-free linear combination of the 90 and 150\,GHz CMB temperature maps ($T_{{90}}(\hat{\textbf{n}})$ and $T_{{150}}(\hat{\textbf{n}})$):
\begin{equation}
\widetilde{T}(\hat{\textbf{n}})  =  \frac{f\widetilde{T}_{{150}}(\hat{\textbf{n}}) - T_{{90}}(\hat{\textbf{n}})}{f-1}
\end{equation}
where
\begin{equation}
\widetilde{T}_{{150}}(\hat{\textbf{n}})  =  T(\hat{\textbf{n}})_{{150}} \ast \frac{B_{{90}}(\hat{\textbf{n}})}{B_{{150}}(\hat{\textbf{n}})},
\end{equation} 
is the 150\,GHz map convolved by the ratio of the 90 and 150\,GHz beam functions. 
The factor $f = 1.67$ is the  ratio  of the tSZ amplitude in 90 and 150 GHz channels.
 We assume the 90\,GHz beam is a $\theta_{\rm FWHM} = 1.'7$ Gaussian, and that the 90 and 150\,GHz maps have equal levels of white noise. 
This linear combination degrades the final lensing significance due to (i) the larger effective beam size and (ii) the higher  map noise level -- for these assumptions tSZ cleaning reduces the SNR by nearly a factor of three. 
In an ideal case, this tSZ cleaning completely removes the tSZ signal and results in no detectable bias (b = $-0.7 \pm 0.7$ percent).
In practice, one is unlikely to know the value of $f$ perfectly due to uncertainties in the relative calibration of the bands, the spectral bandpasses of the bands, or potential relativistic corrections to the tSZ frequency spectrum. 
We approximate this imperfect knowledge of $f$ by assuming a 1\% error in the relative calibration of the 90 and 150\,GHz. 
This level of calibration uncertainty may be overly conservative for future CMB experiments but has already been achieved by current ground-based experiments. The percent-level residual tSZ signal significantly biases the recovered masses, leading the mass to be underestimated by $-6.3\pm0.7$ percent. An under-estimation of the residual tSZ signal can shift the bias direction, just like for the kSZ case.

Instead of using multiple frequencies to subtract the tSZ signal, one might instead use data only from the tSZ null around 217 GHz. 
Unfortunately, the flux from dusty galaxies rises sharply with frequency, so the 217\,GHz maps have substantially more flux from the dusty galaxy members of a cluster than 90 or 150\,GHz maps. 
Recent work has measured the magnitude of the correlation between the tSZ effect and the cosmic infrared background (CIB) due to these dusty galaxies, although  it remains poorly constrained.  
For instance, \citet{george2015} found a tSZ-CIB correlation of  $\eta = 0.113^{ +0.057}_{ -0.054}$; estimating the tSZ-CIB signal at 220\,GHz from Eq.(14) of that work yields a signal of nearly three quarters of the tSZ signal at 150\,GHz (or a power spectrum of order 55\% of the tSZ power spectrum). 
Thus, the CIB signal at 217\,GHz would need to be removed to a comparable level as the tSZ signal. 
%\tbd{reminder: make clear SZ biases can go either way depending on whether over/underestimate them}

\subsubsection{tSZ fitting} 
Instead of removing the tSZ signal, one can include it in the pixel-pixel covariance matrix. 
Following \S\ref{sec_kSZ_bias}, we calculate the expected tSZ contribution using SPH simulations from \citet{mccarthy2013}. 
As would be expected, the bias is consistent with zero for perfect knowledge of the tSZ signal (b = $1.0 \pm 0.6\%$). 
However, this bias increases quickly if the tSZ contribution is mis-estimated. 
Analogously to the tSZ cleaning case, we would expect percent-level errors to lead to significant (order 6\%) biases. 
Given that the current uncertainties in modeling the tSZ signal from galaxy clusters are more than an order of magnitude larger, it will be extremely challenging to achieve the sub-percent precision necessary to make this approach viable.

\begin {table}
%\caption{Systematic bias in the reconstructed lensing mass}
\centering
%\resizebox{0.5\textwidth}{!}{
\begin{tabular}{| l | C{1.5cm}  | C{1.5cm} | C{1.5cm} | C{1.5cm} |}
    \hline
    \multirow{2}{*}{Bias source} & \multicolumn{4}{c|}{Bias \% at $\Delta T = 1.0\ \ukarcmin$}\\
    \cline{2-5}
    & \multicolumn{2}{c|}{Temperature $T_{\rm ML}$} & \multicolumn{2}{c|}{Polarization $QU_{\rm ML}$} \\%\hline
    \cline{2-5}
     & \% bias& error & \% bias & error \\\hline
    tSZ cleaning (1\% residual signal) & -6.3 & 0.50 & 0 & - \\\hline
    kSZ fitting (20\% uncertainty) & -7.3  & 0.31 & 0 & - \\\hline

    DG in the cluster & -4.5 & 1.70 & -1.3 & 1.20\\\hline
    \hline
    Redshift uncertainty & {\ }0.2 & \multirow{2}{*}{0.25} & -0.3 & \multirow{2}{*}{0.90} \\%\hline
    \cline{1-2}\cline{4-4}
    Cluster positions & -7.5 & & -1.5 & \\\hline
    \hline
   Presence of undetectable haloes & 2.5 & 0.25  &6.3 &0.92\\
    \hline\hline
    \multicolumn{5}{|l|}{Uncertainties in cluster mass profile}\\\hline
   ~~~$\kappa_{\rm NFW} + \kappa_{sub}$ & {\ }3.1 & \multirow{4}{*}{0.25} & -0.6 & \multirow{4}{*}{0.92}\\%\hline
    \cline{1-2}\cline{4-4}
    ~~~$\kappa_{Einasto}$ & -2.4 & & -2.5 &\\%\hline
    \cline{1-2}\cline{4-4}
    ~~~$\kappa_{\rm NFW}^{^{mod}}$ & {\ }2.2  & & {\ }0.6 & \\\hline

 \end{tabular}
\vspace*{2mm}
%}
%\tablecomments{
\caption{The percentage mass biases from various systematic uncertainties. A positive (negative) number means the recovered mass is over- (under-)estimated. The first three lines reflect the expected biases from astrophysical foregrounds; these are serious for the temperature estimator, but not the polarization estimator. The next two lines deal with uncertainties in where the cluster is located (whether in redshift or on the sky); these are likely to be manageable for both estimators. The last four lines relate to uncertainties in how the mass is distributed, whether due to projection effects from nearby, lower-mass haloes, or due to uncertainties in the average mass profile for galaxy clusters. These questions about the mass distribution are a concern for both temperature and polarization. }
%\caption{Systematic bias in the reconstructed lensing mass}
\label{tab_sys_bias}
\vspace*{2mm}
\end{table}

\subsection{Dusty galaxies in the cluster and other foregrounds}
\label{sec_DG_sys_bias}

Galaxy clusters are known to host overdensities of dusty galaxies, with several papers measuring the resulting tSZ-CIB correlation \citep{actdunkley2013, george2015, PLANCKTSZCIB2016}. 
We describe our modeling of these DG overdensities\footnote{For implementation reasons, in this section, we include all foregrounds mentioned in the Appendix \ref{sec_appendix}, even ones that are not correlated with the cluster itself, such as radio galaxies.} in the Appendix \ref{sec_appendix_extragal}.
If ignored, the tSZ-CIB correlation may substantially bias the recovered masses from temperature estimators, especially at higher frequencies. 
The emission from dusty galaxies rises sharply with frequency, by an order of magnitude in $\mu K^2_{\rm CMB}$ from 90 to 150\,GHz and again from 150 to 220\,GHz. 
Polarization estimators (at least at 150\,GHz and lower frequencies) are essentially unaffected due to the lower polarization fraction  of dusty galaxies (expected to be less than 4\% \citep{seiffert2007, sptpol_delensing_2017}). 
The tSZ-CIB correlated power could be handled analogously to either the tSZ fitting or cleaning approaches in \S\ref{subsec:tszbias}. 
However, a multi-frequency cleaning scheme will be less effective than for the tSZ effect since the spectral dependence of thermal dust emission varies between individual galaxies. 
Here we look only at bias for the fitting approach where the extra pixel-pixel covariance due to the clustered dusty galaxies is folded into the likelihood. 
The recovered mass is somewhat low: $b=4.5 \pm 1.7\%$. 
The existence of a bias (higher than $2\sigma$) is slightly surprising since one would expect zero bias in the perfect information limit, and the significance is low enough that it may be a statistical fluke. 
The dramatic increase in the uncertainty -- from 0.25\% to 1.7\% -- reflects the plateauing of the dotted line in right panel of Fig.~\ref{fig_delM_M_1000_clusters_T_QU_EB_ideal_FG}. 
Unsubtracted foreground power effectively sets a lower bound on the instrumental noise.

\section{A look into the future}
\label{sec_forecast}

\begin{figure}
\centering
\includegraphics[width=1.\textwidth, height=0.68\textwidth,clip=]{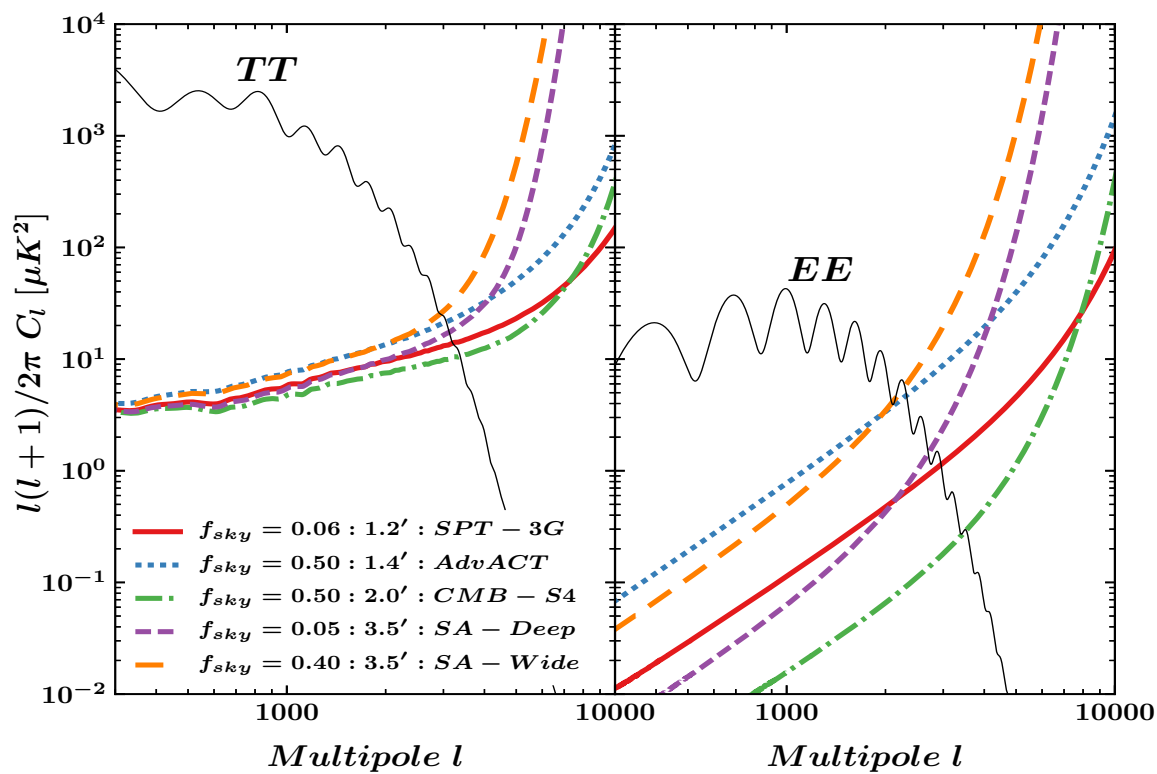}
\caption{The expected residual foreground and noise power spectrum for the future CMB experiments. 
The 90, 150, and 220 GHz channels have been combined using a constrained ILC technique to remove the tSZ effect while minimizing other extragalactic foregrounds and instrumental power. 
The left and the right panels correspond to temperature and polarization respectively.
The plateauing of the residual temperature spectrum reflects the limited foreground removal possible with three frequency channels. 
Specifications about each experiment are listed in Table \ref{tab_forecast_future_CMBexp}.}
\label{fig_ILC_res}
\end{figure}

In this final section we forecast the cluster mass uncertainties from CMB-cluster lensing for the AdvACT, Simons Array, and SPT-3G experiments, which we will collectively refer to as the Stage III experiments, and also for the proposed CMB-S4 experiment. 
In addition to presenting the mass uncertainties for fiducial versions of these experiments, we examine how the mass uncertainty would change as a function of the beam size and map noise levels. 
This information can be used to evaluate design tradeoffs while planning the CMB-S4 experiment. 

\subsection{Expected lensing mass uncertainties for future CMB experiments}
\label{subsec:cmbs3s4}

\begin{table}
%\caption{Forecasted mass uncertainties for the future CMB experiments after foreground cleaning}
\centering
%\small{\resizebox{0.5\textwidth}{!}{
\begin{tabular}{|c | c | c | c | c | c | c | c | c |}
   \hline
    & \multicolumn{3}{c|}{} & \multirow{3}{*}{$f_{sky}$}& Effective & \# of &  & \\
    Experiment & \multicolumn{3}{c|}{$\Delta T$  [\ukarcmin] } & & beam & clusters & $T_{\rm ML}$ & $QU_{\rm ML}$ \\
    \cline{2-4}%\cline{7-8}
    %& { }90 & 150 &  220 & $[\theta_{_{\rm FWHM}}]$  &  & {\ \ }$T_{\rm ML}$  & $QU_{\rm ML}$ \\\hline
    & { }90 & 150 &  220 & & $[\theta_{_{\rm FWHM}}]$  & ($N_{clus}$) & (ILC)  & (ILC) \\\hline
    \multirow{3}{*}{CMB - S4} & \multirow{3}{*}{{ }1.0} & \multirow{3}{*}{{ }1.0} & \multirow{3}{*}{{ }1.0} & \multirow{3}{*}{0.50} & $1.0'$ & \multirow{3}{*}{{ }100,000} & 0.87\% & 0.83\% \\
    \cline{6-6}\cline{8-9}
    & & & &  & $2.0'$ & & 0.95\% & 0.98\% \\
    \cline{6-6}\cline{8-9}
    & & &  &  & $3.5'$ & & 1.20\% & 1.60\% \\\hline
    \hline
    SPT-3G & { }4.5 & { }2.5 & { }4.5 & 0.06 & $1.2'$ &  \multirow{4}{*}{10,000} & 3.28\% &  6.12\% \\%\hline
    \cline{1-5}\cline{6-6}\cline{8-9}
    AdvACT & { }8.0 & { }7.0 & 25.0 & 0.50 & $1.4'$ &  & 4.35\% & $>$15\% \\%\hline
    \cline{1-5}\cline{6-6}\cline{8-9}
   Simons Array - Deep & { }1.5 & { }1.5 & { }4.7 & 0.05 &\multirow{2}{*}{ $3.5'$} &  & 4.41\% & 8.45\% \\%\hline
    \cline{1-5}\cline{8-9}
   Simons Array - Wide & 5.5 & { }5.5 & 20.0 & 0.40 & &  & 5.86\% & $>$15\%\\\hline
 \end{tabular}
%}}
\vspace*{2mm}
%\tablecomments{
\caption{The forecasted mass uncertainties for large-aperture future CMB experiments. We combine data from 90, 150, and 220 GHz to clean the extragalactic foregrounds using a constrained ILC method designed to remove the tSZ signal while minimizing the residual foreground power and instrumental noise.  For polarization, the ILC is essentially optimal weighting of the bands for minimum noise. }
\label{tab_forecast_future_CMBexp}
\end{table}

We expect the next generation of CMB experiments, which will have substantially more detectors and a concomitant reduction in map noise levels, to dramatically improve the cluster mass calibration possible from CMB-cluster lensing. 
The experimental configuration of all the experiments considered is given in Table \ref{tab_forecast_future_CMBexp}. 
Three options for telescope size (and therefore beam sizes) are listed for the proposed CMB-S4 experiment. 
While current results have mass uncertainties of order $\ge 20$\% \citep{baxter2015,act_cmass2015,planckXXIV2015}, we expect Stage III experiments to reach 3\% and CMB-S4 to approach 1\%. 

There are two reasons for the improvements.
First, with more detectors comes lower map noise levels (and larger survey areas).  
The deepest current experiments reach approximately $5\,\mu$K-arcmin in temperature; the Stage III surveys  (AdvACT \citep{advact_2016};  Simons Array \citep{PB2_2016}, and SPT-3G \citep{benson2015_3g}) forecast a few $\mu$K-arcmin; and projections for CMB-S4 are $\sim$\,1\,$\mu$K-arcmin. 
Lower noise improves the lensing significance on any individual galaxy cluster.
Second, lower noise levels and larger survey areas translate into substantially more galaxy clusters. 
Current ground-based SZ cluster catalogs have fewer than 1000 clusters \citep{ACTSZ2013, bleem2015}, but SPT-3G is forecast to find 8000 clusters \citep{benson2015_3g}, AdvACT 10,000 clusters \citep{advact_2016} and we assume ad-hoc that CMB-S4 will find 100,000 clusters. 
In addition to the internally discovered clusters, optical surveys like DES \citep{rykoff2016}, and in the future LSST \citep{lsst_science_book} and Euclid \citep{euclid_science_book}, will yield extremely large numbers of galaxy clusters within the CMB survey regions, as will the X-ray satellite eROSITA \citep{erosita_science_book}. 
This method is perfectly suited to determining the mass calibration for these external cluster catalogs as well. 

\begin{figure*}[tbh]
\centering
\includegraphics[width=0.46\textwidth, height=0.5\textwidth,clip=]{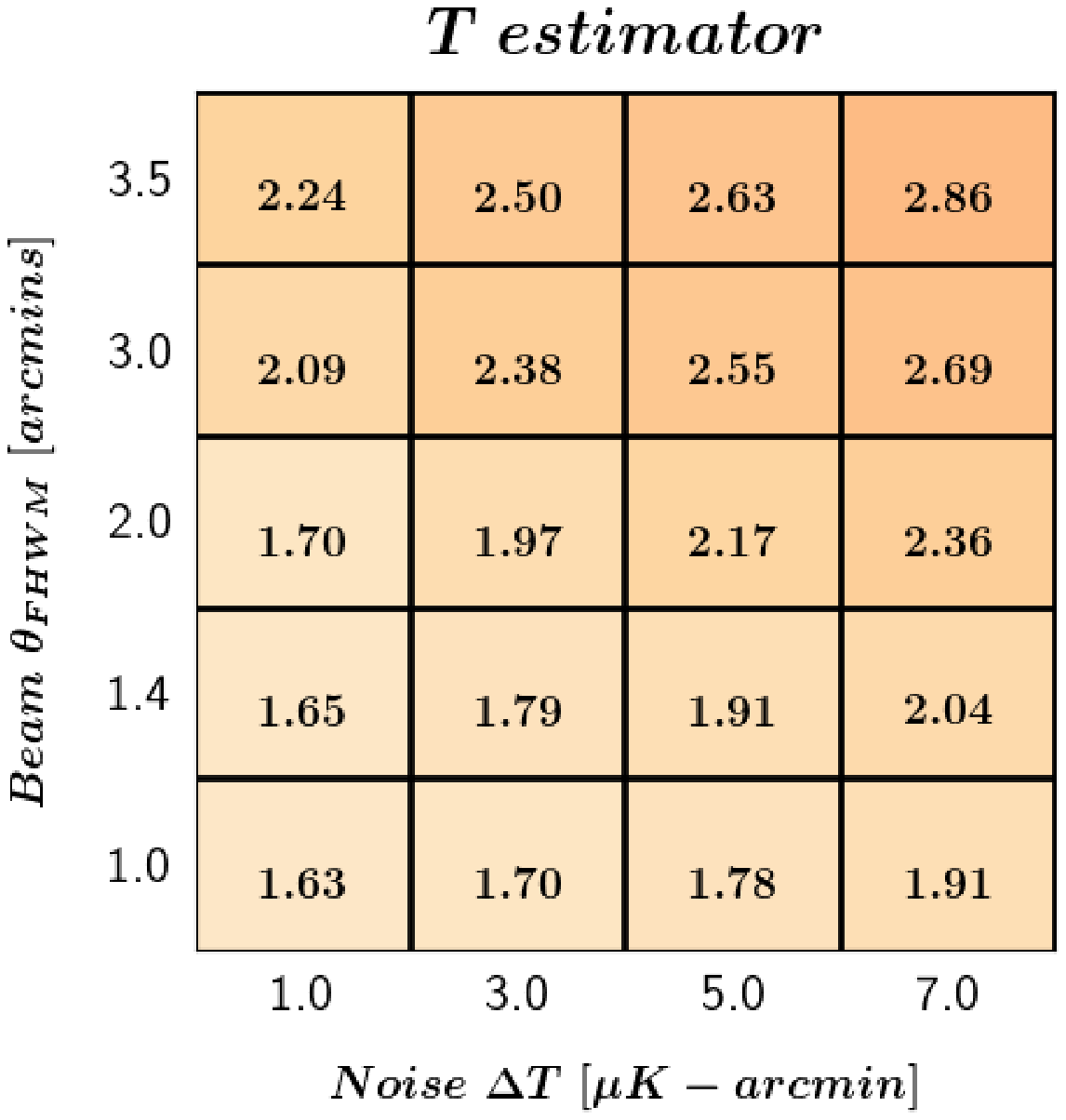}
\includegraphics[width=0.5\textwidth, height=0.5\textwidth,clip=]{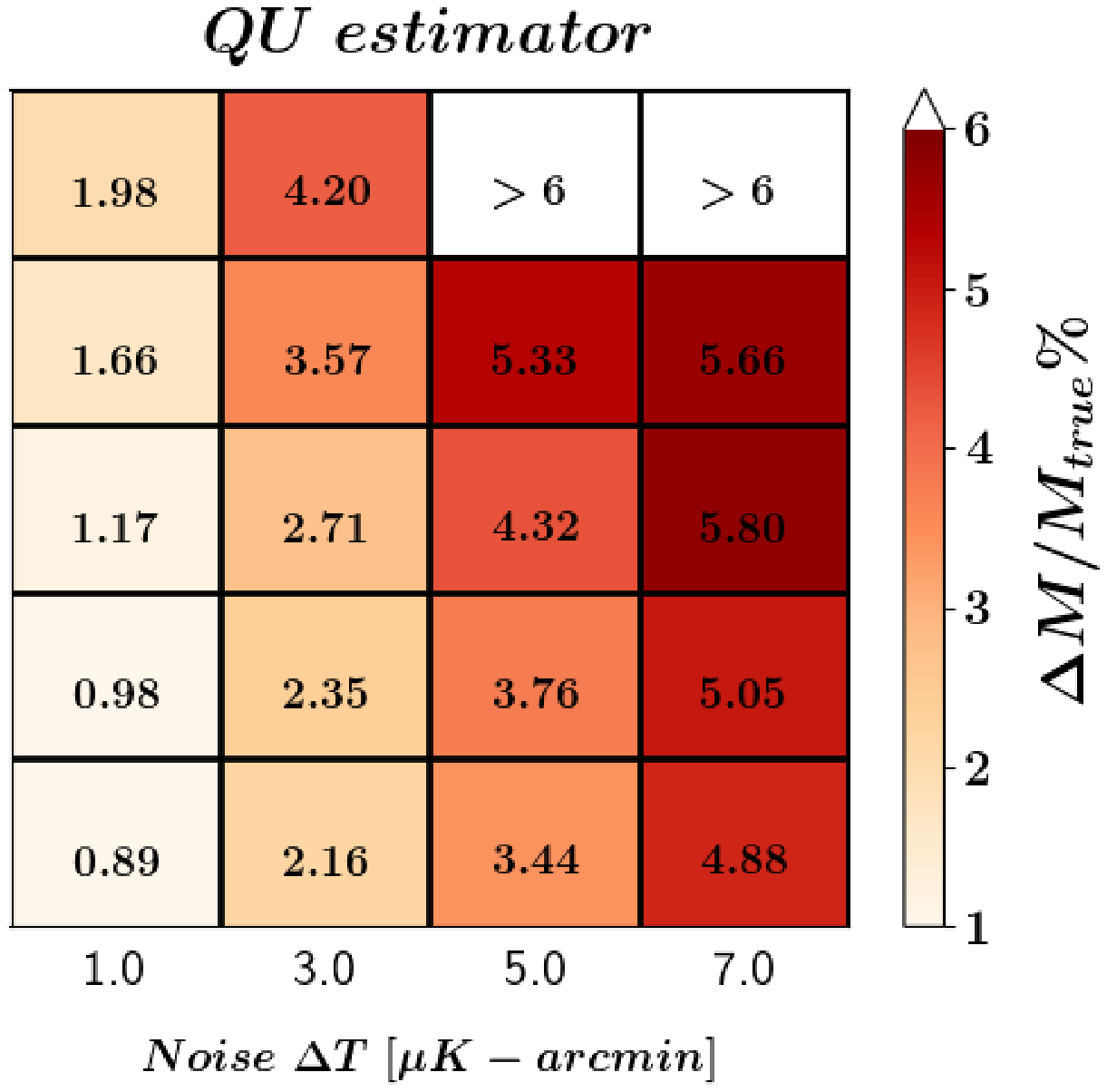}
\caption{The performance of the polarization MLE is very sensitive both to the angular resolution and map noise level of an experiment; the gains for the temperature MLE are much smaller. The numbers correspond to the CMB-cluster lensing mass uncertainty (in percent) of the cluster sample containing 100,000 clusters after the addition of foregrounds (dotted lines in right panel of Fig. \ref{fig_delM_M_1000_clusters_T_QU_EB_ideal_FG}). 
Improving the beam from $3.'5$ to $1'$ enhances the SNR by a factor of two for the CMB-S4 noise levels. 
The saturation of $T_{\rm ML}$ is due to the larger impact of foregrounds on the temperature maps. }
\label{fig_beam_dependency_T_QU}
\end{figure*}

To provide realistic estimates of the mass uncertainties, we perform a constrained internal linear combination (ILC) of data from 90, 150, and 220 GHz channels based on the \texttt{SMICA} (Spectral Matching Independent Component Analysis) algorithm \citep{smicacardoso, PLANCKCOMPSEP2014} to eliminate the tSZ signal from the temperature data and minimize the residual power in other extragalactic foregrounds and instrumental noise in both temperature and polarization. 
The resulting power spectra of the instrumental noise and residual foregrounds for different CMB experiments are shown in Fig.~\ref{fig_ILC_res}. 
At $\ell \le 2000$, the temperature curves are dominated by residual foreground power as three frequency bands are insufficient to completely eliminate the foreground power in the assumed model (see Appendix). 
As a result, the temperature noise curves  converge at $\ell \le 2000$ despite the very different noise levels of the  experiments.  
During this process, we convolve the 90 and 220 GHz spectra by the ratio of 150 GHz beam and their native beams, so that the final effective beam size matches 150 GHz. 

The expected performance of each  experiment is given in Table \ref{tab_forecast_future_CMBexp}. 
One significant uncertainty is the number of clusters to assume for each experiment. 
As accurately modeling the survey selections functions for SZ, optical, and X-ray surveys is beyond the scope of this work, we make the simplifying assumption that all Stage III experiments will have 10,000 clusters and the CMB-S4 experiment will have 100,000 clusters. 
%This is of order the number expected to be discovered through the tSZ effect by the SPT-3G (8000 clusters; \citet{benson2015_3g}) or AdvACT (10,000 clusters; \citet{advact_2016}), but likely an over-estimate for the Simons Array due to its larger 3.5$^\prime$ beam size. 
This is of order the number expected to be discovered through the tSZ effect by the SPT-3G or AdvACT (see above), but likely an over-estimate for the Simons Array due to its larger 3.5$^\prime$ beam size. 
On the other hand, the experimental beam size is irrelevant when predicting the size of cluster samples from optical or X-ray surveys that overlap with the CMB surveys. 
The DES or LSST surveys should provide samples with more than 50,000 clusters for all of the Stage III CMB experiments \citep{rykoff2016, lsst_science_book}. 
Given a specific sample size, the mass uncertainty can be obtained by rescaling the numbers provided in Table \ref{tab_forecast_future_CMBexp} by $\sqrt{\frac{N_{sample}}{N_{clus}}}$.

Even considering concerns about potential biases from astrophysical signals, the temperature channel will be extremely important for the cluster mass estimates from the Stage III CMB experiments.% (AdvACT, Simons Array, SPT-3G). 
The mass uncertainty on the fiducial 10,000 cluster sample  is similar in temperature from all three experiments, with a range from 3.3\% (SPT-3G) to 5.9\% (the Simons Array wide survey). These uncertainties are as large as the likely systematic uncertainties, and the statistical uncertainties on polarization are higher by a factor of two or more. As an example of scaling the results with sample size, we replace the fiducial sample size by the expected number counts for SZ-discovered clusters with SPT-3G (8000) and optically detected clusters from the DES (50,000). SPT-3G would achieve a 3.6\% mass uncertainty with a sample of 8000 SZ-selected clusters and a 1.5\% uncertainty on a sample of 50,000 optically-selected clusters. The shallow portions of the Simons Array or AdvACT surveys cannot contribute much for the polarization estimator; lower noise levels are essential. The polarization estimator can be within a factor of two for the deep surveys of the Simons Array or SPT-3G. For instance, the polarization estimator for SPT-3G on 8000 clusters yields a 6.8\% mass calibration, to be compared to the 3.6\% mass calibration from temperature (ignoring systematic uncertainties).

The lower level of systematic uncertainty for polarization comes into play for the CMB-S4 experiment. First, for the extremely low noise levels of CMB-S4, the performance of the temperature and polarization channels is nearly identical (0.95\% vs.~0.98\%) for an instrument with $2'$ beam resolution. Second, the magnitude of the temperature-only systematic errors (primarily from the SZ effect) is now several times larger than the raw statistical uncertainties, and would dominate the temperature error budget. We can expect cluster mass calibrations from CMB-S4's polarization data at the 1\% level.

The mass calibration forecasts in  Table \ref{tab_forecast_future_CMBexp} are highly complementary  to and competitive with the masses obtained by stacking optical weak lensing measurements. For example, LSST hopes to achieve a mass uncertainty of 1\% by stacking few thousands of clusters at redshifts $z < 0.5$ \citep{lsst_science_book}. At high redshifts, since the number density of background galaxies decrease rapidly, the constraints from optical lensing measurements tend to weaken. Calibrating the high redshift end of the mass function is the true power of CMB-cluster lensing which will allow us to place important constraints on the redshift evolution of mass-observable scaling relations out to high redshifts $z \ge 1.5$.

\subsection{Optimizing survey design: SNR as a function of beam size and map noise levels}
\label{sec_beam_dependence}

There are plans underway to build a substantially more sensitive CMB experiment, CMB-S4 \citep{cmbs4_2016}, with work already underway both on design studies and planning for CMB-S4 and on pathfinder experiments to CMB-S4 such as the Simons Observatory. There is a wide spectrum of science drivers for these experiments, of which CMB-cluster lensing is only one. However, it is useful to consider what CMB-S4 design choices would be optimal for cluster mass calibration. In this section, we consider two lever arms: map noise levels and angular resolution (beam size). %In all cases, we assume three frequency bands centered at 90, 150 and 220\,GHz with equal noise levels and beam sizes that scale as the wavelength. 
We do not consider the number or relative weight of frequency bands, although these decisions will be important for handling astrophysics foregrounds in the temperature estimator. 

In Fig.~\ref{fig_beam_dependency_T_QU}, we present the mass uncertainties on a sample of 100,000 clusters from the temperature and polarization estimators for a grid of five different beam sizes $\theta_{{\rm FWHM}} = 1^\prime, 1.^\prime4, 2^\prime, 3^\prime, 3.^\prime 5$, and four temperature map noise levels (1, 3, 5, or 7\,\ukarcmin{}). We have simplified the problem by using only 150\,GHz data with no foreground removal. As a result, the quoted uncertainties are likely to be too large for both temperature and polarization estimators. 
However, the qualitative conclusions are robust to this assumption as will be shown below by spot-checking the results with a full ILC analysis.

Notably, the temperature results show only a minor improvement ($<20$\%) going from 7 to 1\,\ukarcmin{} noise levels. The plateauing occurs because the instrumental noise is already smaller than the foreground power. Although the exact level may be off, these 150 GHz only results are consistent with the picture from the full, 3-frequency ILC analysis shown in Fig.~\ref{fig_ILC_res}. In that figure, the residual foreground and noise power curves for temperature are essentially the same, whether from CMB-S4, from small and deep Stage III surveys, or from wide and shallow Stage III surveys. In short, foreground residual power dominates the results even at the lower sensitivities of the Stage III experiments. 
The temperature estimator also shows a fairly modest effect from reducing the beam size: a factor of 3.5 reduction in beam size from $3.^\prime5$ to $1^\prime$ only improves the mass uncertainty by a factor of 1.35. This is because the foreground power floor   limits the use of small-scale modes where the beam size matters most. 

In contrast, as seen in the right panel of Fig.~\ref{fig_beam_dependency_T_QU}, the fidelity with which the QU polarization estimator recovers cluster masses is strongly dependent on the experimental noise level and beam size. The clear improvements are because polarization estimator is still noise instead of residual foreground dominated. Improving map noise levels from 3 to 1\,\ukarcmin{} improves the SNR on CMB-cluster lensing by a factor of 2.4 (the equivalent change for the temperature estimator is only 1.04). Similarly, reducing the beam size threefold from $3^\prime$ to $1^\prime$ leads to an improvement by a factor of 1.9. Both the beam size and instrumental noise levels matter for the performance of the polarized CMB-cluster lensing estimator. 

Finally, we confirm this picture by extending these 150\,GHz-only predictions to 3-band data using the ILC method. We assume equal temperature map noise levels of $1 \mu$K-arcmin at 90, 150 and 220\,GHz, with a beam size that scales with the wavelength. We consider three 150\,GHz beam sizes, $1^\prime, 2^\prime$, and $3.^\prime 5$, with the results  tabulated in Table~\ref{tab_forecast_future_CMBexp}. 
The improvement  as a function of the beam size is consistent between the 150\,GHz and ILC cases. While the improvement is only marginal for the temperature channel, the mass uncertainty from the polarization estimator drops to 0.83\% for a $1^\prime$ beam, a factor of 1.9 better than the results for a $3.^\prime5$ beam. The corresponding improvement for 150\,GHz only is slightly better, at a factor of 2.2. 

\section{Conclusion}
\label{sec_conclusion}
We have developed MLEs to optimally extract lensing information from the temperature (T) and polarization (Q/U) maps of the CMB. We show a Q/U based MLE recovers as much information as an estimator using E- and B-mode maps. We also show that the temperature MLE performs better than the standard QE by a factor of two at very low noise levels in the absence of astrophysical foregrounds; the performance gain is not significant for the polarization estimator due to the lower SNR. We consider the effects of these foregrounds on the cluster lensing estimators, finding at 150\,GHz that  astrophysical foregrounds have no impact on the polarization MLE and set an effective noise floor of a few \ukarcmin{} on the temperature MLE unless removed using multi-frequency data. 
 
We quantify the systematic uncertainties due to astrophysical foregrounds (the tSZ effect, kSZ effect or dusty galaxies), uncertainty in the cluster position or redshift, projection effects from nearby lower-mass haloes, and uncertainty in the cluster mass profile. We find the astrophysical foregrounds are likely to significantly bias the temperature MLE at the 4.5 - 7.3\% level; the polarization MLE is largely unaffected. The biases due to uncertainty in the cluster position or redshift are manageable for both temperature and polarization. Lower mass haloes near the galaxy cluster lead to an overestimate of the cluster masses at the 2.5 to 6.3\% level, and will need to be carefully accounted for using simulations. 
The uncertainties in the cluster mass profile can shift the cluster mass either up or down by up to a few percent. Better measurements of the cluster mass profile are needed to reduce this uncertainty.

Finally, we present forecasts for the mass uncertainties from upcoming CMB experiments,  combining multiple frequency bands with an ILC technique to minimize the instrumental noise and astrophysical foregrounds. The AdvACT, Simons Array and SPT-3G experiments will achieve mass calibration uncertainties of order 3 - 6\% for a sample containing 10,000 clusters, with the temperature channel being crucial to these mass constraints. With the even lower noise levels of CMB-S4 and a $2^\prime$ beam, we find the statistical mass uncertainty from either the temperature or polarization MLEs falls to just below 1\% with 100,000 clusters. We expect polarization to be the main information channel for CMB-S4 given the potential biases  due to the temperature foregrounds. Finally, we consider how the performance of CMB-S4 depends on the assumed noise level or beam size, finding that a factor of three reduction in either the beam size or noise level leads to roughly a factor of two improvement on the mass calibration from the polarization MLE. CMB-S4 has the potential to transform galaxy cluster cosmology by reducing the current 20\% mass uncertainty on galaxy clusters twentyfold to $\sim$\,1\%. 

\section*{Acknowledgments}
We thank Kimmy Wu for revising the paper draft and Nathan Whitehorn for useful discussions. We thank the anonymous referee for some useful suggestions that helped in shaping this manuscript better. We thank the high performance computation centre at University of Melbourne for providing access to the cluster \texttt{spartan.unimelb.edu.au}. We acknowledge the use of \texttt{HEALPix} \citep{healpix} and \texttt{CAMB} \citep{lewis2000} routines. This work was performed in the context of the South Pole Telescope scientific program. SPT is supported by the National Science Foundation through grant PLR-1248097.  Partial support is also provided by the NSF Physics Frontier Center grant PHY-1125897 to the Kavli Institute of Cosmological Physics at the University of Chicago, the Kavli Foundation and the Gordon and Betty Moore Foundation grant GBMF 947. The work at the University of Melbourne is supported by the Australian Research Council's Discovery Projects scheme (DP150103208). LB's work was supported under the U.S. Department of Energy contract DE-AC02-06CH11357.

\appendix
\section{Simulated skies}
\label{sec_appendix}
The MLE presented in this work depends being able to produce large numbers of realistic simulated skies, incorporating a diverse range of astrophysical signals: the CMB (lensed by the galaxy cluster), radio galaxies, dusty galaxies, the kSZ effect, and the tSZ effect. 
One unique challenge for cluster studies is that the galaxy cluster itself can source most of these signals in addition to contributions from other unrelated haloes. These simulated skies are used for the calculation of the pixel-pixel covariance matrix, and for the creation of mock data sets (\S\ref{sec_pixel_pixel_cov_matrix}). In this appendix, we detail the creation of these these sky simulations. 

The sequence of operations is as follows. First, simulations of each signal are created on $50^\prime \times 50^\prime$ boxes with a $0.25^\prime$ pixel resolution. Most of this appendix will focus on how this is done. The CMB maps are then lensed by the galaxy cluster, convolved by a Gaussian beam of the appropriate size, and rebinned to $0.5^\prime$ pixels. This final rebinning reduces the number of map pixels four-fold and substantially speeds up the MLE without significantly reducing the SNR. White, Gaussian instrumental noise is added, with a pixel RMS level based on an experiment's sensitivity, observing time and survey area. For computational reasons, the $50^\prime \times 50^\prime$ box is cut down to the central $10^\prime \times 10^\prime$ area used in the analysis. Finally we point out a subtle effect because of constraining the simulations to a $50^\prime \times 50^\prime$ box. Choosing a smaller box will reduce the background CMB gradient and subsequently the lensing signal generated by the cluster -- which will tend to worsen $\Delta M/M$ --. To quantify this, we repeated our simulations with a larger $2^{\circ} \times 2^{\circ}$ such that it encompasses the first peak of the CMB. At $\Delta T = 1 \mu K-arcmin$ for a sample of 100,000 clusters, the mass uncertainty $\Delta M/M$ is now 0.237\% as opposed to 0.252\% in the left panel of Fig \ref{fig_delM_M_1000_clusters_T_QU_EB_ideal_FG} -- a very small effect. 

Since we are dealing with very small areas of sky, we adopt the flat-sky approximation and substitute Fourier transforms for spherical harmonic transforms. The Fourier wavenumber $k$ is related to the multipole $\ell$ by $k = \sqrt{k_x^{2} + k_y^{2}} = \frac{\ell}{2\pi}$. We define the azimuthal angle $\phi_{\ell}$ as $tan^{-1}(k_{y} / k_x)$.

\subsection{Cosmic Microwave Background}
\label{sec_appendix_CMB}

To simulate CMB maps that have been lensed by a massive galaxy cluster, we begin by creating T, Q, and U maps that are Gaussian realizations \citep{kamionkowski1996} of the CMB power spectra ($C_{\ell}^{TT}, C_{\ell}^{TE}, C_{\ell}^{EE},$ and $C_{\ell}^{BB}$). 
\iffalse{Note that the Q and U spectra are set by $C_{\ell}^{EE}$ and $C_{\ell}^{BB}$ according to:
\begin{eqnarray}
Q_{\ell} &=& E_{\ell}\ cos (2 \phi_{\ell}) - B_{\ell}\ sin (2 \phi_{\ell})\\
U_{\ell} &=& E_{\ell}\ sin (2 \phi_{\ell}) + B_{\ell}\ cos (2 \phi_{\ell})
\label{eq_QU_from_EB}
\end{eqnarray}}\fi
For these fiducial power spectra, we use the lensed CMB power spectra predicted by \texttt{CAMB} \citep{lewis2000} for the \emph{Planck} 2015 $\Lambda$CDM cosmology\footnote{More specifically, we use the best-fit parameters from the \emph{Planck} 2015 chain that combines the \emph{Planck} 2015 temperature, polarization, lensing power spectra with BAO, $H_{0}$, and SNe data (\texttt{TT,TE,EE+lowP+lensing+ext} in Table 4 of \cite{planckcosmo2015}).}. Note that the tensor-to-scalar ratio $r$ is zero in this chain, and there is no contribution from inflationary B-modes \citep{baumann2009}. By using Gaussian realizations of the lensed CMB power spectra, we are effectively assuming (1) that the lensing due to large-scale structures (LSS) occurs at higher redshift than the galaxy cluster, and (2) that the small non-Gaussianities due to this LSS lensing are negligible. We then lens the T, Q, and U maps using with the cluster convergence profile described in \S\ref{sec_cluster_profile}. 
We deal with sub-pixel deflection angles by interpolating over the maps using a fifth-degree B-spline interpolation. 

\subsection{Sunyaev-Zel'dovich (SZ) effect}
\label{sec_appendix_SZ}

There are two SZ effects of interest: the kinematic SZ (kSZ) effect and the thermal SZ (tSZ) effect. Both SZ effects have contributions from the cluster itself as well as from unrelated haloes. For the latter signal, we assume the best-fit tSZ and kSZ power spectra from \citet{george2015}. 
We use the \citet{shaw2010} model for the tSZ power spectrum with an overall normalization of \mbox{$D_{\ell=3000}^{^{\rm tSZ, 150 GHz}} = 3.7\ \mu K^{2}$}. For the kSZ spectrum, we take the \citet{shaw2012} model  with an overall normalization of \mbox{$D_{\ell=3000}^{^{\rm kSZ}} = 2.9\, \mu K^{2}$}. For the uncorrelated tSZ and kSZ signals, we simply create Gaussian realizations of these two spectra. 

In some cases, we wish to study the effect of the kSZ and tSZ signals from the galaxy cluster in question. %At these times, we replace the simple Gaussian realizations by cutouts around comparable mass haloes in simulated SZ maps. 
At these times, we add the cutouts around comparable mass haloes in simulated SZ maps to the simple Gaussian CMB realizations. We use the kSZ maps provided for the \citet{flender2016} N-body simulations, while for tSZ maps, we use the smoothed-particle hydrodynamics (SPH) simulations of  \citet{mccarthy2013}. 

We ignore the extremely small polarized SZ signals in all cases. To generate polarization the free electrons of the intracluster medium must be exposed to a quadrupole radiation field, due to the CMB quadrupole mode for tSZ polarization, $p_{tSZ}$, and to an apparent CMB quadrupole created by the Doppler effect of bulk velocities in the electrons for kSZ polarization, $p_{kSZ}$. The level of the tSZ polarization is $p_{tSZ} \sim 0.1 (\tau_{e}/0.02)\ \mu K$ while the kSZ polarization level is $p_{kSZ} \sim 0.1 \beta_{t}^{2}\tau_{e}\ K$ \citep{sazonov1999, carlstrom2002} where $\tau_{e}$ is the optical depth of the cluster and $\beta_{t} = v/c$ transverse component of the electron's velocity. The clusters used in this work have $M_{200} =  2 \times 10^{14}\ M_{\odot}$ and an expected optical depth of $\tau_{e} \sim 0.004$ \citep{flender2016b}. 
We assume a velocity, $v = 1000\ km\ s^{-1}$, leading to $p_{tSZ}=20$\,nK and $p_{kSZ} =2$\,nK. This level of polarization is negligible.

\subsection{Radio and Dusty Galaxies}
\label{sec_appendix_extragal}

We also create simulated maps of radio and dusty galaxies. 
These maps include four terms: 
\begin{enumerate}
\item[1.] radio galaxies following a spatial Poisson distribution ($C_{\ell}^{^{radio}} \propto {\rm constant}$),
\item[2.] dusty star forming galaxies (DGs) also following a spatial Poisson distribution ($C_{\ell}^{^{\rm DG-Po}} \propto {\rm constant}$),
\item[3.] clustering of DGs ($C_{\ell}^{^{\rm DG-clus}} \propto  \ell^{0.8}$), and 
\item[4.] the overdensity of DGs in galaxy clusters
\end{enumerate} 
The first three items are uncorrelated with the galaxy cluster and are expected to reduce the SNR of the estimators without biasing the reconstructed lensing mass. However, the relative overdensity of DGs in galaxy clusters can potentially bias the derived masses, and is discussed  in \S\ref{sec_sys_bias_checks}. We assume the galaxies are randomly polarized at the 4\% level, based on \citet{sptpol_delensing_2017}. This 4\% level is expected to be an over-estimate for the DGs \citep{seiffert2007}. 

For radio galaxies, we draw a Poisson realization of the number counts as a function of flux \citep{dezotti2005}. We ignore the clustering of radio galaxies as it is irrelevant at frequencies observed by the CMB telescopes \citep{dezotti2005, george2015}. We take a more sophisticated approach for the dusty galaxies to handle the clustering and tSZ-CIB correlation. We begin by taking a Poisson distribution for the unusually bright dusty galaxies ($>1$\,mJy) \citep{bethermin2011} . For fainter DGs, we create a set of number density contrast maps $\delta(\hat{\textbf{n}})^{^{\rm DG-Po}}$ covering narrow flux bins. We adjust these number density contrast maps to account for the desired clustering and correlation properties as outlined below, and then assign random fluxes to each pixel in Jy/sr (drawn uniformly across the flux bin). The resulting flux maps are then converted to CMB temperature units ($\ \mu K_{\rm CMB}$). 

For the tSZ-CIB correlation, we use the tSZ simulations produced by \cite{mccarthy2013} and the $C_{\ell}^{^{\rm tSZ \times CIB}}$ cross spectrum measured by \citet{george2015}. Using these we produce a $\rm tSZ \times CIB$ correlated map $T(\hat{\textbf{n}})^{^{\rm tSZ \times CIB}}$ and a weight map $W(\hat{\textbf{n}})^{^{\rm DG \times tSZ}}$ to modify $\delta(\hat{\textbf{n}})^{^{\rm DG-Po}}$ as
\begin{eqnarray}
T_{\ell} \equiv T_{\ell}^{^{\rm tSZ \times CIB}} & = & T_{\ell}^{^{\rm tSZ}}\ \frac{C_{\ell}^{^{\rm tSZ \times CIB}}}{C_{\ell}^{^{\rm tSZ}}}\\
\label{eq_DG_tSZ_weight_map}
W(\hat{\textbf{n}}) \equiv W(\hat{\textbf{n}})^{^{\rm DG \times tSZ}} & = & \frac{\left| T(\hat{\textbf{n}}) \right|}{\sum{\left|T(\hat{\textbf{n}})\right|}}\\
\label{eq_DG_tSZ_density_map}
\widetilde{\delta}(\hat{\textbf{n}})^{^{\rm DG \times tSZ}} & = & W(\hat{\textbf{n}})\ \delta(\hat{\textbf{n}})^{^{\rm DG-Po}}
\end{eqnarray} where subscripts $\ell$ refer to the harmonic space transforms of the CMB map $T(\hat{\textbf{n}})$. Eq.(\ref{eq_DG_tSZ_weight_map}) and Eq.(\ref{eq_DG_tSZ_density_map}) ensure that the number of point sources are conserved and are predominantly clustered near massive dark matter haloes. For the clustering component of DG we modify Eq.(\ref{eq_DG_tSZ_density_map}) following \cite{gonzalez2004}
\begin{eqnarray}
\widetilde{\delta}_{\ell}^{^{DG}} \equiv \widetilde{\delta}_{\ell}^{^{\rm DG-Po, clus, tSZ}} & = & \widetilde{\delta}_{\ell}^{^{\rm DG \times tSZ}}\ \frac{ \sqrt{ C_{\ell}^{^{Po}} + C_{\ell}^{^{clus}}} } { \sqrt{ C_{\ell}^{^{Po}} } }
\end{eqnarray}

\vspace*{5mm}

%%%%%%%%%%%%% Bibliography %%%%%%%%%%%%%

%\bibliographystyle{}
{}

\end{document}